\documentclass{autart}

\usepackage{natbib}
\usepackage{IEEEtrantools}

\usepackage{cite}
\usepackage{booktabs} 
\usepackage{array} 
\usepackage{paralist} 
\usepackage{verbatim} 
\usepackage{subfig} 
\usepackage{mathtools}
\usepackage{mathrsfs}
\usepackage{amsfonts}
\usepackage{amsmath}
\usepackage{amssymb}
\usepackage{bbm}     

\newcommand{\ie}{i.e.,~}
\newcommand{\eg}{e.g.,~}

\newcommand{\T}{\mathcal{T}}

\renewcommand{\Psi}{\varPsi}
\renewcommand{\Gamma}{\varGamma}
\renewcommand{\Lambda}{\varLambda}
\renewcommand{\Phi}{\varPhi}
\renewcommand{\Omega}{\varOmega}
\renewcommand{\Sigma}{\varSigma}
\renewcommand{\Theta}{\varTheta}
\renewcommand{\Pi}{\varPi}
\renewcommand{\Upsilon}{\varUpsilon}

\newcommand{\beq}{\begin{equation}}
\newcommand{\eeq}{\end{equation}}
\newcommand{\beqn}{\begin{equation*}}
\newcommand{\eeqn}{\end{equation*}}
\newcommand{\beqarr}[1]{\begin{IEEEeqnarray}{#1}}
\newcommand{\eeqarr}{\end{IEEEeqnarray}}
\newcommand{\beqarrn}[1]{\begin{IEEEeqnarray*}{#1}}
\newcommand{\eeqarrn}{\end{IEEEeqnarray*}}
\newcommand{\bbmat}{\begin{bmatrix}}
\newcommand{\ebmat}{\end{bmatrix}}

\newcommand{\kronecker}{\raisebox{1pt}{\ensuremath{\:\otimes\:}}}  

\newcommand{\norm}[1]{\left \Vert #1 \right \Vert}

\newcommand{\wint}[1]{\int_{-\pi}^{\pi} \!\!\!  #1 \: \mathrm{d} \omega}
\newcommand{\expect}[1]{\mathrm{E}\left [ #1 \right]}

\newcommand{\var}[1]{\mathrm{Var} \: #1 }

\newcommand{\defeq}{\vcentcolon=} 

\newcommand{\tr}[1]{\operatorname{\text{Tr}} \left\lbrace #1 \right\rbrace}

\newcommand{\diag}[1]{\mathrm{diag}\left\lbrace #1 \right\rbrace}

\newcommand{\vect}{\mathrm{vec}}

\newcommand{\bnul}{\begin{enumerate}[a)]}
\newcommand{\enul}{\end{enumerate}}

\newtheorem{proposition}{Proposition}[section]

\newtheorem{lemma}{Lemma}[section]
\newtheorem{assumption}{Assumption}[section]

\newtheorem{example}{Example}[section]

\newcounter{counter}

\newcommand{\Abf}{{\mathbf A}}

\newcommand{\Kbf}{{\mathbf K}}

\newcommand{\Pbf}{{\mathbf P}}

\newcommand{\Rbf}{{\mathbf R}}
\newcommand{\Sbf}{{\mathbf S}}

\newcommand{\Wbf}{{\mathbf W}}

\newcommand{\bbf}{{\mathbf b}}

\newcommand{\ebf}{{\mathbf e}}

\newcommand{\gbf}{{\mathbf g}}

\newcommand{\sbf}{{\mathbf s}}

\newcommand{\wbf}{{\mathbf w}}

\newcommand{\zbf}{{\mathbf z}}

\newcommand{\Rb}{{\mathbb R}}

\usepackage{tikz} 
\usepackage{pgfplots}
\pgfplotsset{compat=newest}
\usetikzlibrary{calc,intersections,through,backgrounds,shapes.misc,positioning,chains,scopes}
\tikzset{subsystem/.style={
rectangle,
minimum size=6mm,
thick,
draw=black,
top color=white, 
bottom color=white, 
font=\itshape,
text height=1.5ex,
text depth=0.25ex
}}

\tikzset{operator/.style={
circle,minimum size=15pt,
inner sep=0pt,
thin,draw=black, 
top color=white,bottom color=white,
text height=1.75ex,
text depth=0.25ex,
font=\itshape}}

\tikzset{signal/.style={
circle,minimum size=9pt,
inner sep=0pt,
thin,draw=black, 
top color=white,bottom color=white,
text height=1.5ex,
text depth=0.25ex,
font=\itshape}}

\tikzset{point/.style={circle,inner sep=0pt,minimum size=0pt,fill=black}}
\tikzset{skip loop/.style={to path={-- ++(0,#1) -| (\tikztotarget)}}}

\usepackage{caption}
\usepackage{pgfplots}
\usepgfplotslibrary{statistics}
\pgfplotsset{
  compat=1.12,
    boxplot/draw direction=y,
    boxplot/every box/.style={solid,black},
    boxplot/every outlier/.style={/tikz/mark=*,},
    boxplot/every whisker/.style={black,solid},
    boxplot/draw/whisker/.code 2 args={%
    \draw[/pgfplots/boxplot/every whisker/.try,/tikz/dashed]
    (boxplot cs:#1) -- (boxplot cs:#2);
    \draw[/pgfplots/boxplot/every whisker/.try]
    (boxplot whisker cs:#2,0) -- (boxplot whisker cs:#2,1);
    },%
    boxplot/whisker extend={\pgfkeysvalueof{/pgfplots/boxplot/box extend}*0.4},%
}
\usepackage{pgfplotstable,booktabs}
\usepackage{blockdiagram}
\usepackage{color}
\usepackage{nicefrac}

\definecolor{red}{RGB}{157,16,45}
\definecolor{light-red}{RGB}{228,54,62}
\definecolor{green}{RGB}{98,146,46}
\definecolor{light-green}{RGB}{176,201,43}
\definecolor{blue}{RGB}{25,84,166}
\definecolor{light-blue}{RGB}{36,160,216}
\definecolor{pink}{RGB}{216,84,151}
\definecolor{yellow}{RGB}{250,185,25}
\definecolor{dark-grey}{RGB}{101,101,108}
\definecolor{grey}{RGB}{189,188,188}
\definecolor{light-grey}{RGB}{227,229,227}
\usetikzlibrary{calc,matrix,intersections,through,backgrounds,shapes.misc,positioning,chains,scopes}

\DeclareMathOperator*{\argmin}{arg\,min}
\DeclareMathOperator*{\argmax}{arg\,max}

\newlength\figureheight
\newlength\figurewidth
\begin{document}

\begin{frontmatter}
\title{An empirical Bayes approach to identification of modules in dynamic networks}

\thanks[footnoteinfo]{This work was  supported by the Swedish Research Council under contracts 2015-05285 and 2016-06079, and by the European Research Council under the advanced grant LEARN, contract 267381.
}

\author[kth]{Niklas Everitt}\ead{neveritt@kth.se},
\author[kth]{Giulio Bottegal}\ead{bottegal@kth.se},
\author[kth]{H\aa kan Hjalmarsson}\ead{hjalmars@kth.se}

\address[kth]{ACCESS Linneaus Center, School of Electrical Engineering, KTH Royal Institute of Technology, Sweden}


\begin{abstract}
We present a new method of identifying a specific module in a
dynamic network, possibly with feedback loops. Assuming known topology,
we express the dynamics by an acyclic network composed of two blocks
where the first block accounts for the relation between the known
reference signals and the input to the target module, while the second block
contains the target module. Using an empirical Bayes approach, we model the first
block as a Gaussian vector with covariance matrix (kernel) given by the recently
introduced stable spline kernel. The parameters of the target module are estimated
by solving a marginal likelihood problem with a novel iterative scheme based on
the Expectation-Maximization algorithm. Additionally, we extend the method to
include additional measurements downstream of the target module. Using
Markov Chain Monte Carlo techniques, it is shown that the same iterative scheme
can solve also this formulation. Numerical experiments illustrate the
effectiveness of the proposed methods.
\end{abstract}
\end{frontmatter}

\section{Introduction}
Networks of dynamical systems are everywhere, and applications are in
different branches of science, e.g., econometrics, systems biology, social
science, and power systems. Identification of these networks, usually referred to
as \emph{dynamic networks}, has been given increasing attention in the system
identification community, see \eg \citet{materassi2010topological}, \citet{VandenHof2013},
\citet{Hjalmarsson2009}.

In this paper, we use dynamic network to mean the interconnection of
\emph{modules}, where each module is a linear time-invariant (LTI) system.
The interconnecting signals are the outputs of these modules.
In a graph interpretation, the interconnecting signals represent nodes and the
modules represent the edges of the graph. Moreover, we assume that exogenous
measurable signals may affect the dynamics of the network.

Two main problems arise in dynamic network identification. The first is
unraveling the network topology (i.e., identify the edges of the graph), which
can be seen as a model structure selection problem. The second problems is the
identification of one or more specific modules in the network.

Some recent papers deal with both the aforementioned problems
\citep{materassi2012problem,chiuso2012bayesian,materassi2010topological,Hayden2014networkMinPhase},
whereas others are mainly focused on the identification of a single module in the
network \citep{dankers2013predictor,Gunes2014,dankers2015errors,Haber2014,Torres2014}.
In particular, \citet{dankers2013predictor}, and \citet{VandenHof2013} study the
problem of understanding which of the available output measurements should be
used to obtain a consistent estimate of a target module. In
\citet{dankers2015errors} instead, errors-in-variables dynamic networks are
considered, and methods that lead to consistent module estimates are proposed.
As observed in \citet{VandenHof2013}, dynamic networks with known topology can
be seen as a generalization of simple compositions, such as systems in cascade,
series or feedback connection. Therefore, identification techniques for dynamic
networks may be derived by extending methods already developed for simple structures.
This is the idea underlying the method presented in \citet{VandenHof2013}, which
generalizes the two-stage method, originally developed for closed-loop systems,
to dynamic networks \citep{Forssell1999}. Instrumental variable methods for
closed-loop systems \citep{Gilson2005} are adapted to networks in
\citet{dankers2015errors}.
Similarly, the methodology proposed in \citet{wahlberg2009variance} for the
identification of cascaded systems is generalized to the context of dynamic
networks in \citet{Gunes2014}. In that work, the underlying idea is that a
dynamic network can be transformed into an acyclic structure, where any
reference signal of the network is the input to a cascaded system consisting of
two LTI blocks.
In this alternative system description, the first block captures the relation
between the reference and the noisy input of the target module, the second block
contains the target module. The two LTI blocks are identified simultaneously
using the prediction error method (PEM) \citep{ljung1998system}. In this setup,
determining the model structure of the first block of the cascaded structure may
be complicated, due to the possibly large number of interconnections in the
dynamic network. Furthermore, it requires knowledge of the model structure of
essentially all modules in the feedback loop. Therefore, in \citet{Gunes2014},
the first block is modeled by an unstructured finite impulse response (FIR)
model of high order. The major drawback of this approach is that, as is usually
the case with estimated models of high order, the variance of the estimated FIR
model is high. The uncertainty in the estimate of the FIR model of the first
block will in turn decrease the accuracy of the estimated target module.

The objective of this paper is to propose a method for the identification of a
module in dynamic networks that circumvents the high variance that is due to the high
order model of the first block. The main contributions of this paper are two-fold.
First, we discuss the case where only the sensors directly measuring the input
and the output of the target module are used in the identification process.
Following a recent trend in system
identification, we use regularization to control the variance
\citep{chen2012estimation}.
In particular, by exploiting the equivalence between regularization and Gaussian
process regression \citep{pillonetto2014kernel}, we model the impulse response
of the first block as a zero-mean stochastic process. The covariance matrix is
given by the recently introduced \emph{first-order stable spline kernel}
\citep{pillonetto2010new}, whose structure is parametrized by two \emph{hyperparameters}.
An estimate of the target module is then obtained by empirical Bayes (EB)
arguments, that is, by maximization of the marginal likelihood of the available
measurements \citep{pillonetto2014kernel}.
This likelihood depends not only  on the parameter of the target module, but
also on the kernel hyperparameters and the variance of the measurement noise.
Therefore, it is required to estimate all these quantities. This is done by
designing a novel iterative solution scheme based on an EM-type algorithm
\citep{dempster1977maximum}, known as the Expectation/Conditional-Maximization
(ECM) algorithm \citep{meng1993maximum}, which alternates the so called expectation
step (E-step) with a series of conditional-maximization steps (CM-steps).
When only the module input and output sensors are used, the E-step admits an
analytical expression, because joint likelihood of the module output and the
sensitivity function is Gaussian. As for the CM-steps, one has to solve
relatively simple optimization problems, which either admit a closed form solution,
or can be efficiently solved
using gradient descent strategies. Therefore, the overall optimization scheme for
solving the marginal likelihood problem turns out computationally efficient.

The second main contribution of the paper deals with the case where more sensors
spread in the network are used in the identification of the target module.
Adding information through addition of measurements used in the identification
process has the potential to further reduce the variance of the estimated
module \citep{everitt2017variance}. The downside is that an additional
measurement comes with another module to estimate, also increasing the
number of parameters to estimate. To keep the  number of additional parameters
to estimate low, we propose a method that exploits regularization, modeling as
a Gaussian process also the impulse response of the path linking the target
module to any additional sensor. In this case, however, the measured outputs and
the unknown paths do not admit a joint Gaussian description. As a consequence,
the E-step of the ECM method does not admit an analytical expression, as opposed
to the one-sensor case described above. To overcome this issue, we use
Markov Chain Monte Carlo (MCMC) techniques \citep{gilks1995markov} to solve the
integral associated with the E-step. In particular, we design an integration
scheme based on the Gibbs sampler \citep{geman1984stochastic} that, in
combination with the ECM method, builds up a novel identification method for the
target module reminiscent of the so called empirical Bayes Gibbs sampling
\citep{casella2001empirical}.

The effectiveness of the proposed methods is demonstrated through numerical
experiments. The methods proposed in this paper are close in spirit to some
recently proposed kernel-based techniques for blind system identification
\citep{bottegal2015blind} and Hammerstein system identification
\citep{risuleo2015kernel}. A part of this paper has previously been presented in
\citet{everitt2016identification}. More specifically, the case where only the
sensors directly measuring the input and the output of the target module are used
in the identification process where partly covered in
\citet{everitt2016identification}, whereas, the method where more sensors
spread in the network are used in the identification of the target module is
completely novel.

The paper is organized as follows. In the next section, we introduce the dynamic
network model and we give the problem statement. In Section
\ref{sec:EB_estimation} we present the identification strategy. In Section
\ref{sec:marginal}, we describe the solution scheme based on the ECM algorithm.
Additional measurements are added in Section~\ref{sec:extension}, and we present
the MCMC based scheme to estimate the target module.
Section~\ref{sec:experiments} reports the results of Monte Carlo experiments.
Some conclusions end the paper.


\subsection{Notation} Given a sequence of scalars $\{a(t)\}_{t=1}^m$, we denote
by $a$ its vector representation $a = [a(1) \,\cdots \,a(m)]^T \in \Re^{m}$.
Given a vector $a\in \Re^m$, we define by $\T_n(a)$
the $m\times n$ lower triangular Toeplitz matrix whose elements are the entries of $a$.
Lower case letters indicate, in general, column vectors and, when there is no
confusion, capital letters indicate their Toeplitz form, so given $a\in \Re^m$,
we have that $A~=~T_n(a)$, where the number $n$ of columns is consistent with
the rest of the formula. The symbol ``$\otimes$'' denotes the standard Kronecker
product of two matrices.

\section{Problem Statement} \label{sec:problem_statement}

\subsection{Dynamic networks}

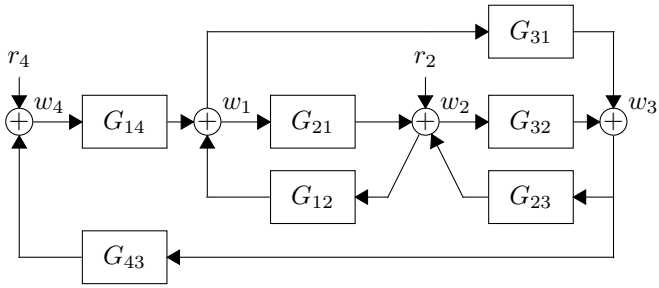
\begin{figure}[!ht]
    \centering
  \begin{tikzpicture}[scale=1,>=triangle 60,tip/.style={->,shorten >=-0.5pt},every join/.style={rounded corners,shorten >=-0.6pt}] 
\matrix[row sep={6mm,between origins},column sep={5mm,between origins}] {
		& [9mm]	& [6mm]	\node (p13) [point] {};	& [9mm]	& [5mm]  & 
		[0mm]	\node (p16) [point] {}; &   &  [4mm]
\node (G24) [block,minimum width=3.2em] {$G_{31}$};	& [6mm]
	\node (p19) [point] {};
\\    
\node (p21) [point,label={[yshift=-0.0mm,xshift=0mm]$r_4$}] {}; & & & & & \node (p26) [point,label={[yshift=-0.0mm,xshift=0mm]$r_2$}] {};  & & 
\\
\node (w1) [sum,label={[yshift=-2mm,xshift=4mm]$w_4$}] {$+$}; &
\node (G21) [block,minimum width=3.2em] {$G_{14}$}; & 
\node (w2) [sum,label={[yshift=-2mm,xshift=4mm]$w_1$}] {$+$};  &
\node (G32) [block,minimum width=3.2em] {$G_{21}$}; & &
\node (w3) [sum,label={[yshift=-2mm,xshift=4mm]$w_2$}] {$+$};  & &
\node (G43) [block,minimum width=3.2em] {$G_{32}$}; & 
\node (w4) [sum,label={[yshift=-2mm,xshift=4mm]$w_3$}] {$+$}; 
\\
[4mm]
& & \node (p43) [point] {};  &
\node (G23) [block,minimum width=3.2em] {$G_{12}$}; & 
\node (p45) [point] {};  & &
\node (p47) [point] {};  &
\node (G34) [block,minimum width=3.2em] {$G_{23}$};
& \node (p49) [point] {}; 
\\
[2mm]
\node (p51) [point] {};					&	
\node (G14) [block,minimum width=3.2em] {$G_{43}$};	&
& & & & &  & \node (p59) [point] {};
\\
};

{ [start chain]
	\chainin (w1);
	\chainin (G21)		[join=by tip];
	\chainin (w2) 		[join=by tip];
	\chainin (G32) 		[join=by tip];
	\chainin (w3)		[join=by tip];
	\chainin (G43)		[join=by tip];
	\chainin (w4)		[join=by tip];
	\chainin (p49)		[join];
	\chainin (p59)		[join];
	\chainin (G14)		[join=by tip];
	\chainin (p51)		[join];
	\chainin (w1)		[join=by tip];
}
{ [start chain]
	\chainin (p49) 		[join=by tip];
	\chainin (G34) 		[join=by tip];
	\chainin (p47)		[join];
	\chainin (w3)		[join=by tip];
	\chainin (p45)		[join];
	\chainin (G23) 		[join=by tip];	
	\chainin (p43)		[join];
	\chainin (w2)		[join=by tip];
}
{ [start chain]
	\chainin (w2) 		[join];
	\chainin (p13) 		[join];
	\chainin (G24) 		[join=by tip];
	\chainin (p19)		[join];
	\chainin (w4)		[join=by tip];
}
{ [start chain]
	\chainin (p21);
	\chainin (w1)		[join=by tip];
}
{ [start chain]
	\chainin (p26);
	\chainin (w3)		[join=by tip];
}

\end{tikzpicture}
  \caption{Network example of 4 internal variables and 2 reference signals.}
   \label{fig:network}
\end{figure}

\begin{figure}[!ht]
    \centering


\begin{tikzpicture}[scale=1,>=triangle 60,tip/.style={->,shorten >=-0.5pt},every join/.style={rounded corners,shorten >=-0.6pt}]
\matrix[matrix of math nodes,nodes in empty cells,draw] (s) at (0,2)
{
 S_{12}(q) 	& 	S_{14}(q)	\\
 S_{22}(q) 	& 	S_{24}(q) 	\\
};
\matrix[matrix of math nodes,nodes in empty cells,draw,minimum width=30pt,above right=-0.57cm and 1.7cm of s] (g1) 
{
   \\ \\
};
\node at (g1) {$G_{31}(q)$};
\matrix[matrix of math nodes,nodes in empty cells,draw,minimum width=30pt,below right=-0.58cm and 1.7cm of s] (g2)
{
   \\ \\
};
\node at (g2) {$G_{32}(q)$};

\node (wj) [sum,below right=-0.85 and 3.5 cm of s,label={[yshift=-4.5mm,xshift=5mm]$w_3$}] {$+$};

\node (r1) [above left=-0.53 and 0.8cm of s,label={[yshift=0.4mm,xshift=2mm]$r_2$}] {};
\node (r2) [below left=-0.52 and 0.8cm of s,label={[yshift=-2mm,xshift=2mm]$r_4$}] {};

\node (w1) [sum,above right=-0.34 and 0.7cm of s,label={[yshift=-2mm,xshift=4mm]$w_1$}] {$+$} ;
\node (w2) [sum,below right=-0.35 and 0.7cm of s,label={[yshift=-2mm,xshift=4mm]$w_2$}] {$+$};

\draw[->] (r1)  |- (s.162);
\draw[->] (r2)  |- (s.198);
\draw[->] (s.+21) |- (w1) ;
\draw[->] (s.-21.2) |- (w2) ;

{ [start chain]
	\chainin (w1);
	\chainin (g1) 		[join=by tip];
}
{ [start chain]
	\chainin (w2);
	\chainin (g2) 		[join=by tip];
}

\draw[->] (g1.+1) -| (wj) ;
\draw[->] (g2.-1) -| (wj) ;

\end{tikzpicture}

  \caption{Direct acyclic graph of part of the network in Figure~\ref{fig:network}.}
   \label{fig:network_sensitivity}
\end{figure}
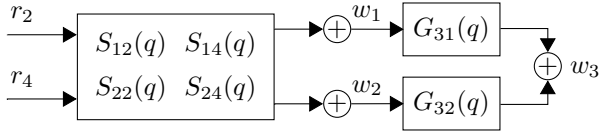

We consider dynamic networks that consist of $L$ scalar \emph{internal variables}
$w_j(t)$, $j = 1,\dotsc,L$ and $L$ scalar external \emph{reference signals}
$r_l(t)$, $l = 1,\dotsc,L$, that can be manipulated by the user.
Some of the reference signal may not be present, i.e., they may be identically zero.
Define $\mathcal{R}$ as the set of indices of reference signals that are present.
In the dynamic network, the internal variables are considered nodes and transfer
functions are the edges. Introducing the vector notation
$w(t):=[w_1(t)\,\ldots\,w_L(t)]^T$, $r(t):=[r_1(t)\,\ldots\,r_L(t)]^T$, the
dynamics of the network are defined by the equation
\begin{IEEEeqnarray*}{rCl}
w(t)
&=&
\mathcal{G}(q)w(t) + r(t) \,,
\yesnumber
\label{eq:network_dynamics}
\end{IEEEeqnarray*}
where
\begin{IEEEeqnarray*}{rCl}
\mathcal{G}(q)
&=&
\begin{bmatrix}
0 & G_{12}(q) & \cdots & G_{1L}(q) \\
G_{21}(q) & 0 & \ddots & \vdots \\
\vdots & \ddots & \ddots & G_{(L-1)L}(q) \\
G_{L1}(q) &  \cdots & G_{L(L-1)}(q) & 0
\end{bmatrix}  \,,
\end{IEEEeqnarray*}
where $G_{ji}(q)$ is a proper rational transfer function for $j=1, \dotsc,L$, $i=1, \dotsc, L$.
The internal variables $w(t)$ are measured with additive white noise, that is
\begin{IEEEeqnarray*}{rCl}
\tilde w(t) & = & w(t) + e(t) \,,
\end{IEEEeqnarray*}
where $e(t) \in \Rb^L $ is a stationary zero-mean Gaussian white-noise process with diagonal noise covariance matrix $\Sigma_e = \diag{\sigma^2_1,\dotsc,\sigma^2_L}$. We assume that the $\sigma_i^2$ are unknown.
To ensure stability and causality of the network the following assumptions hold for all networks considered in this paper.
\begin{assumption}
The network is well posed in the sense that all principal minors of $\lim_{q \to
\infty}(I-\mathcal{G}(q))$ are non-zero \citep{VandenHof2013}.
\end{assumption}
\begin{assumption}
The sensitivity path $S(q)$
\begin{IEEEeqnarray*}{rCl}
S(q) & \defeq & (I-\mathcal{G}(q))^{-1} \,
\end{IEEEeqnarray*}
is stable.
\end{assumption}
\begin{assumption}
The reference variables $\{r_l(t)\}$ are mutually uncorrelated and uncorrelated with the measurement noise $e(t)$.
\end{assumption}
Thus, we can write
\begin{IEEEeqnarray}{rCl}
\tilde w(t) & = & S(q)r(t) + e(t)\,.
\label{eq:Closed_loop}
\end{IEEEeqnarray}
We define a $\mathcal{N}_j$ as the set of indices of internal variables that
have a direct causal connection to $w_j$, \ie $i \in \mathcal{N}_j$ if and only
if $G_{ji}(q) \neq 0 $. Without loss of generality, we assume that
$\mathcal{N}_j = 1,2,\dotsc, p$, where $p$ is the number of direct causal
connections to $w_j$ (we may always rename the nodes so that this holds).
The goal is to identify module $G_{j1}(q)$ given $N$ measurements of the
reference $r(t)$, the ``output'' $\tilde w_j(t)$ and the set of $p$ neighbor
signals in $\mathcal{N}_j$.
To this end, we express $\tilde w_j$, the measured output of module $G_{j1}(q)$ as
\begin{IEEEeqnarray*}{rCl}
\tilde w_j(t) & = & \sum_{i \in \mathcal{N}_j} G_{ji}(q)w_i(t) + r_j(t) + e_j(t) \,.
\label{eq:w_from_G} \yesnumber
\end{IEEEeqnarray*}
The above equation depends on the internal variables $w_i(t)$, $i \in \mathcal{N}_j$,
which we we only have noisy measurement of; these can be expressed as
\begin{equation}
\tilde w_i(t) = w_i(t) + e_i(t) = \sum_{l \in \mathcal{R}} S_{il}(q)r_l(t) + e_i(t) \,.
\label{eq:w_from_S}
\end{equation}
where $S_{il}(q)$ is the transfer function path from reference $r_l(t)$ to
output $\tilde w_i(t)$. Together, \eqref{eq:w_from_G} and \eqref{eq:w_from_S}
allow us to express the relevant part of the network, possibly containing
feedback loops, as a direct acyclic graph with two blocks connected in cascade.
Note that, in general, the first block depends on all other blocks in the network.
Therefore, accurate low order parametrization of this block depends on global
knowledge of the network.
\begin{example} \label{expl:network}
As an example consider the network depicted in Figure~\ref{fig:network}, where,
using \eqref{eq:w_from_G} and \eqref{eq:w_from_S}, the acyclic graph of
Figure~\ref{fig:network_sensitivity} can describe the relevant dynamics,
when $w_j=w_3$ is the output and we wish to identify $G_{31}(q)$.

\end{example}

In the following, we briefly review two standard methods for closed-loop
identification that we will use as a starting point to derive the methodology
described in the paper.

\subsection{A two stage method}

The first stage of the two-stage method \citep{VandenHof2013}, proceeds by finding a consistent estimate $\hat w_i(t)$ of all nodes $w_i(t)$ in $\mathcal{N}_j$.
This is done by high-order modeling of $\{ S_{il} \}$ and estimating it from \eqref{eq:w_from_S} using the prediction error method.
The prediction errors are constructed as
\begin{IEEEeqnarray*}{rCl}
\varepsilon_i(t,\alpha) & = & \tilde w_i(t) - \sum_{l \in \mathcal{R}} S_{il}(q,\alpha)r_l(t) \,,
\label{eq:epsilon_first_stage} \yesnumber
\end{IEEEeqnarray*}
where $\alpha$ is a parameter vector. The resulting estimate $S_{il}(q,\hat \alpha)$ is then used to obtain the node estimate as
\begin{IEEEeqnarray*}{rCl}
\hat w_i(t) & = & \sum_{l \in \mathcal{R}} S_{il}(q,\hat \alpha)r_l(t) \,.
\label{eq:w_hat_first_stage} \yesnumber
\end{IEEEeqnarray*}
In a second stage, the module of interest $G_{j1}(q)$ (and the other modules in $\mathcal{N}_j$) is parameterized by $\theta$ and estimated from \eqref{eq:w_from_G}, again using the prediction error method. The prediction errors are now constructed as
\begin{IEEEeqnarray*}{rCl}
\varepsilon_j(t,\theta) & = & \tilde w_j(t) - r_j(t)- \sum_{i \in \mathcal{N}_j} G_{ji}(q,\theta)\hat w_i(t) \,.
\label{eq:epsilon_second_stage_two} \yesnumber
\end{IEEEeqnarray*}

\subsection{Simultaneous minimization of prediction errors} \label{sec:SMPE}

It is useful to briefly introduce the simultaneous minimization of prediction
error method (SMPE) \citep{Gunes2014}.
The main idea underlying SMPE
is that if, the two prediction errors \eqref{eq:epsilon_first_stage} and
\eqref{eq:epsilon_second_stage_two} are simultaneously minimized, the variance
will be decreased \citep{wahlberg2009variance}.
In the SMPE method, the prediction error of the measurement $\tilde w_j$ depends explicitly on $\alpha$ and is given by
\begin{IEEEeqnarray*}{rCl}
\varepsilon_j(t,\theta,\alpha) & = & \tilde w_j(t) - \sum_{i \in \mathcal{N}_j} G_{ji}(q,\theta) \sum_{l \in \mathcal{R}} S_{il}(q,\alpha)r_l(t) \,.
\label{eq:epsilon_second_stage} \yesnumber \IEEEeqnarraynumspace
\end{IEEEeqnarray*}
The method proceeds to minimize
\begin{equation}
V_N(\theta,\alpha) = \frac{1}{N} \sum_{t=1}^N \left [ \frac{\varepsilon_j^2(t,\theta,\alpha)}{\sigma^2_j} +  \sum_{i \in \mathcal{N}_j} \frac{\varepsilon_i^2(t,\alpha)}{\sigma^2_i}   \right] \,.  \label{eq:V_SMPE}
\end{equation}
In \citep{Gunes2014}, the noise variances are assumed known, and how to estimate the noise variances is not analyzed.
As an initial estimate of the parameters $\theta$ and $\alpha$, the minimizers of the two-stage method can be taken.

The main drawback is that the least-squares estimation of $S$ may still induce high variance in the estimates.
Additionally, if each of the $n_s$ estimated transfer functions in $S$ is estimated by the first $n$ impulse response coefficients, the number of estimated parameters in $S$ alone is $n_s \cdot n$.
Already for relatively small dimensions of $S$ the SMPE method is prohibitively expensive.
To handle this, a frequency domain approach is taken in \citet{Dankers2015nonparametric}.
In this paper, we will instead use regularization to reduce the variance and the complexity.

\section{Empirical Bayes estimation of the module} \label{sec:EB_estimation}
In this section we derive our approach to the identification of a specific module based on EB.
For ease of exposition, we give a detailed derivation in the one-reference-one-module case.
The extension to general dynamic networks follows along similar arguments.

\begin{figure}
    \centering
  \begin{tikzpicture}[scale=1,>=triangle 60,tip/.style={->,shorten >=-0.5pt},every join/.style={rounded corners,shorten >=-0.6pt}] 
\matrix[row sep={6mm,between origins},column sep={12mm,between origins}] {
\node (p11) [point,label={$r_1$}] {};
  & \node (S11) [block,minimum width=3.2em] {$S_{11}(q)$};
  & \node (w1) [sum,label={[yshift=-1mm,xshift=3mm]$w_1$}] {$+$};
  & \node (G) [block,minimum width=3.2em] {$G_{21}(q)$};
  & \node (w2) [sum,label={[yshift=-1mm,xshift=3mm]$w_2$}] {$+$}; \\
};

{ [start chain]
  \chainin (p11);
  \chainin (S11)  [join=by tip];
  \chainin (w1)   [join=by tip];
  \chainin (G)    [join=by tip];
  \chainin (w2)   [join=by tip];
}
\end{tikzpicture}
  \caption{Basic network of 1 reference signal and 2 internal variables.}
   \label{fig:neb-network}
\end{figure}
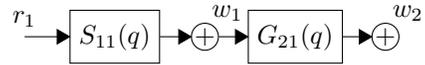

We consider a dynamic network with one non-zero reference signal $r_1(t)$.
Without loss of generality, we assume that the module of interest is $G_{21}(q)$,
and hence $G_{22}(q), \dotsc, G_{2L}(q)$ are assumed zero
(We can always rename the signals such that this holds).
The setting we consider has been illustrated in Figure~\ref{fig:neb-network}.
We parametrize the target module by means of a parameter vector
$\theta \in \mathbb R^{n_\theta}$. Using the vector notation introduced in the
previous section, we denote by $\tilde w_1$ the stacked measurements
$\tilde w_{1}(t)$ before the module of interest $G_{21}(q,\theta)$, and by
$\tilde w_2$ the stacked output of this module $\tilde w_2(t)$.
We define the impulse response coefficients of $G_{21}(q,\theta)$ by
the inverse discrete-time Fourier transform
\begin{equation}
g_\theta(t) \coloneqq \frac{1}{2\pi} \wint{G_{21}(e^{j\omega},\theta)e^{j\omega t}} \,.
\end{equation}
Similarly we define $s_{11}$ as the impulse response coefficients of $S_{11}(q)$,
where $S_{11}(q)$ is, as before, the sensitivity path from $r_1(t)$ to $w_1(t)$,
and $e_1(t)$ and $e_2(t)$ are the measurement noise sources
(which we have assumed white and Gaussian).
Their variance is denoted by $\sigma^2_{1}$ and $\sigma^2_{2}$, respectively.
We rewrite the dynamics as
\begin{equation}
\left.
\begin{IEEEeqnarraybox}[][c]{rCrCr}
\tilde w_1 & = & S_{11} r_1 + e_1 \,,
\\
\tilde w_2 & = & G_\theta S_{11} r_1 + e_2 \,,
\end{IEEEeqnarraybox}
\right.
\label{eq:dynamcs_2}
\end{equation}
where $G_{\theta}$ is the $N \times N$ lower triangular Toeplitz matrix
of the $N$ first impulse response samples $g_\theta$.
The same notation holds for the impulse response $s_{11}$ and its Toeplitz-matrix
version $S_{11} = \mathcal{T}_N(s_{11})$.
We further rewrite \eqref{eq:dynamcs_2} as
\begin{align}
\left.
\begin{IEEEeqnarraybox}[][c]{rCrCl}
\tilde w_1 & = & R_1 s_{11} & + & e_1 \,, \\
\tilde w_2 & = & G_\theta R_1 s_{11} & + & e_2 \,.
\end{IEEEeqnarraybox}
\right.
\label{eq:dynamcs_3}
\end{align}
where $R_1 = \mathcal{T}_N(r_1)$.
For computational purposes, we only consider the first $n$ samples of $s_{11}$,
where $n$ is large enough such that the truncation captures the dynamics of the
sensitivity $S_{11}(q)$ well enough.
Let $z:=[\tilde w_1^T \,\tilde w_2^T]^T$; we rewrite \eqref{eq:dynamcs_3} as
\begin{IEEEeqnarray}{rCl+rCl+rCl}
z & = & W_\theta s_{11} + e \,, & W_\theta & = &
\begin{bmatrix}
R_1 \\
G_\theta R_1
\end{bmatrix}
&
e & = &
\begin{bmatrix}
e_1 \\
e_2
\end{bmatrix}
\IEEEeqnarraynumspace
\end{IEEEeqnarray}
Note that $e$ is a random vector such that
\begin{equation}
\Sigma_e := \expect{ee^T } = \begin{bmatrix}
\sigma^2_{1} I & 0 \\
0 & \sigma_2^2 I
\end{bmatrix} \,.
\end{equation}



\subsection{Bayesian model of the sensitivity path}
To reduce the variance in the sensitivity estimate
(and also reduce the number of estimated parameters), we cast our problem in a
Bayesian framework and  model the sensitivity function as a zero-mean Gaussian
stochastic vector \citep{Rasmussen}, \ie
\begin{IEEEeqnarray}{rCl} \label{prior_s}
p(s_{11};\lambda, K_\beta) \sim \mathcal{N} (0, \lambda K_{\beta}) \,.
\IEEEeqnarraynumspace
\end{IEEEeqnarray}
The structure of the covariance matrix is given by the \emph{first-order stable
spline kernel} \citep{pillonetto2010new}:
\begin{IEEEeqnarray}{rCl+rCl}
\{K_{\beta}\}_{i,j} & = & \beta^{\max(i,j)} , & \quad \beta & \in & [0,\,1) \,.
\IEEEeqnarraynumspace \label{eq:stablespline}
  \end{IEEEeqnarray}
The parameter $\beta$ regulates the decay velocity of the realizations from \eqref{prior_s}, whereas, $\lambda$ tunes their amplitude. In this context, $K_\beta$ is usually called a {\it kernel} (due to the connection between Gaussian process regression and the theory of reproducing kernel Hilbert space, see e.g. \citet{Rasmussen} for details) and determines the properties of the realizations of $s$. In particular, the stable spline kernel enforces smooth and BIBO stable realizations  \citep{pillonetto2010new}.
\subsection{The marginal likelihood estimator}
Since $s_{11}$ is assumed stochastic, it admits a probabilistic description jointly with the vector of observations $z$, parametrized by the vector
\begin{IEEEeqnarray}{rCl}
\eta &=& \begin{bmatrix}
\sigma_1^2 & \sigma_2^2 & \lambda & \beta & \theta
\end{bmatrix} \,.
\IEEEeqnarraynumspace \label{eq:jointly_gaussian_description}
\end{IEEEeqnarray}
In particular, having assumed a Gaussian distribution of the noise, the joint description is also Gaussian, that is,
\begin{IEEEeqnarray}{rCl}
p\left(\begin{bmatrix}
z \\ s_{11}
\end{bmatrix}
; \eta \right) \sim \mathcal{N} \left(
\begin{bmatrix}
0 \\ 0
\end{bmatrix}
,
\begin{bmatrix}
\Sigma_z & \Sigma_{zs} \\
\Sigma_{sz} & \lambda K_\beta
\end{bmatrix}
\right) \,,
\IEEEeqnarraynumspace
\end{IEEEeqnarray}
where
$\Sigma_z = W_\theta\lambda K_\beta W_\theta^T + \Sigma_e$, and $\Sigma_{zs} = \Sigma_{sz}^T =  W_\theta \lambda K_\beta$.
It is instrumental to derive the posterior distribution of $s_{11}$ given the measurement vector $z$.
It is given by \citep{Anderson:1979}
\begin{IEEEeqnarray}{rCl} \label{eq:posterior_s}
p(s_{11}|z;\,\eta) &\sim & \mathcal{N} (
PW_\theta^T\Sigma_e^{-1}z, P) \,,  \\
P &=& (W_\theta^T\Sigma^{-1}_e W_\theta + (\lambda K_\beta)^{-1})^{-1} \,,
\IEEEeqnarraynumspace
\end{IEEEeqnarray}
and it is also parametrized by the vector $\eta$.

The module identification strategy we propose in this paper relies on an
empirical Bayes approach.
We introduce the marginal probability density function (pdf) of the measurements
\begin{equation}
p(z;\, \eta) = \int p(z,\,s_{11})\, ds_{11} \sim \mathcal{N}(0,\,\Sigma_z) \,,
\end{equation}
that is, the pdf of the measurements after having integrating out the dependence on the sensitivity path $s_{11}$. Then, we can define the (log) marginal likelihood (ML) criterion as the maximum of the marginal pdf defined above
\begin{align} \label{eq:marg_lik}
\hat \eta  & = \argmax_{\eta} p(z;\, \eta) \\
& = \argmin_{\eta} \left( \log \det \Sigma_z + z^T \Sigma_z^{-1} z \right) \,,   \nonumber
\end{align}
whose solution provides also an estimate of $\theta$ and thus of the module of interest.

\section{Computation of the solution of the marginal likelihood criterion} \label{sec:marginal}
Problem \eqref{eq:marg_lik} is nonlinear and may involve a large number of decision variables, if $n_\theta$ is large.
In this section, we derive an iterative solution scheme based on the
Expectation/Conditional-Maximization (ECM) algorithm \citep{meng1993maximum}, which is a generalization of the standard Expectation-Maximization (EM) algorithm.
In order to employ EM-type algorithms, one has to define a \emph{latent variable}; in our problem, a natural choice is $s_{11}$.
Then, a (local) solution to \eqref{eq:marg_lik} is achieved by iterating over the following steps:
\begin{enumerate}
\item[(E-step)] Given an estimate $\hat \eta^{(k)}$ (computed at the $k$-th iteration of the algorithm), compute
    \begin{equation} \label{eq:estep}
    Q^{(k)}(\eta) \defeq \mathbb{E} \left[\log p(z,\,s_{11};\,\eta) \right] \,,
    \end{equation}
    where the expectation is taken with respect to the posterior of $s_{11}$ when the estimate $\eta^{(k)}$ is used, i.e., $p(s_{11}|z,\,\hat \eta^{(k)})$\,;
\item[(M-step)] Solve the problem
    \begin{equation} \label{eq:Q_function}
    \hat \eta^{(k+1)} = \argmax_{\eta}  Q^{(k)}(\eta) \,.
    \end{equation}
\end{enumerate}
First, we turn our attention on the computation of the E-step, \ie the derivation of \eqref{eq:estep}.
Let $\hat s_{11}^{(k)}$ and $\hat P^{(k)}$ be the posterior mean and covariance matrix of $s_{11}$, computed from \eqref{eq:posterior_s} using $\hat \eta^{(k)}$.
Define $\hat S_{11}^{(k)} \defeq \hat P^{(k)} + \hat s_{11}^{(k)}\hat s_{11}^{(k)T}$. The following proposition
provides an expression for the function $Q^{(k)}(\eta)$.
\begin{lemma} \label{th:Q_function}
Let $\hat \eta^{(k)} = [\hat \sigma_1^{2(k)} \, \hat \sigma_2^{2(k)} \, \hat \lambda^{(k)} \, \hat \beta^{(k)} \, \hat \theta^{(k)}]$ be an estimate of $ \eta$ after the $k$-th iteration of the EM method.
Then
\begin{equation} \label{eq:Q_decomposition}
 Q^{(k)}(\eta) = -\frac{1}{2} Q_0^{(k)}(\sigma^2_1,\,\sigma^2_2,\,\theta) -\frac{1}{2} Q_s^{(k)}(\lambda,\,\beta) \,,
\end{equation}
where
\begin{IEEEeqnarray}{rCl}
Q_0^{(k)}(\sigma_1^2,\,\sigma_2^2,\,\theta)
	& = & \left( \log \det \Sigma_e + z^T \Sigma_e^{-1}z
    	-2z^TW_{\theta} \hat s_{11}^{(k)}
        \vphantom{\tr{W_{\theta}^T \Sigma_e^{-1} W_{\theta} \hat S_{11}^{(k)} }}
        \right. \nonumber \\
    &  &  \left.  + \tr{W_{\theta}^T \Sigma_e^{-1} W_{\theta} \hat S_{11}^{(k)} } \right) \,,  \!\!\!\!\!\!  \label{eq:Q_0}  \\
Q_s^{(k)}(\lambda,\,\beta)
	& = &\log \det \lambda K_\beta  +  \tr{ \left(\lambda
    	K_\beta\right)^{-1} \hat S_{11}^{(k)} } \,. \nonumber \\
\IEEEeqnarraynumspace \label{eq:Q_s}
\end{IEEEeqnarray}
\end{lemma}
Having computed the function $Q^{(k)}(\eta)$, we now focus on its maximization.
We first note that the decomposition \eqref{eq:Q_decomposition} shows that the kernel hyperparameters can be updated independently of the rest of the parameters:
\begin{proposition} \label{th:update_hps}
Define
\begin{equation} \label{eq:Q_beta}
 Q_{\beta}(\beta) =  \log\det {K_\beta} + n\log \tr{K_{\beta}^{-1}\hat S_{11}^{(k)}}.
\end{equation}
Then
\begin{align}
\hat \beta^{(k+1)} &=  \argmin_{\beta\in[0,1)}Q_{\beta}(\beta)\label{eq:thm_beta} \,, \\
\hat \lambda^{(k+1)} &= \frac{1}{n} \tr{K_{\hat \beta^{(k+1)}}^{-1}\hat S_{11}^{(k)}} \,. \label{eq:thm_lambda}
\end{align}
\end{proposition}
Therefore, the update of the scaling hyperparameter is available in closed-form, while the update of $\beta$ requires the solution of a scalar optimization problem in the domain $[0,\,1]$, an operation that requires little computational effort, see \citet{Bottegal2016} for details.

We are left with the maximization of the function $Q_0^{(k)}(\sigma_1^2,\,\sigma_2^2,\,\theta)$.
In order to  simplify this step, we split the optimization problem into constrained subproblems that involve fewer decision variables.
This operation is justified by the ECM paradigm, which, under mild conditions
\citep{meng1993maximum}, guarantees the same convergence properties of the EM algorithm even when the optimization of $Q^{(k)}(\eta)$ is split into a series of constrained subproblems.
In our case, we decouple the update of the noise variances from the update of $\theta$.
By means of the ECM paradigm, we split the maximization of $ Q_0^{(k)}(\sigma^2_1,\,\sigma^2_2,\,\theta)$ in a sequence of two constrained optimization subproblems:
\begin{align}
    \hat \theta^{(k+1)}   & = \argmax_{\theta} {Q}_0^{(k)}(\sigma^2_1,\,\sigma^2_2,\,\theta) \label{eq:CM_steps1}\\
          &\qquad \mbox{s.t. } \sigma^2_1 = \hat\sigma^{2(k)}_1 ,\, \sigma^2_2 = \hat\sigma^{2(k)}_2 \,, \nonumber\\
    \hat \sigma^{2(k+1)}_1,\,\hat\sigma^{2(k+1)}_2 & = \argmax_{\sigma^{2}_1,\,\sigma^{2}_2} {Q}_0^{(k)}(\sigma^2_1,\,\sigma^2_1,\,\theta) \label{eq:CM_steps2}\\
          &\qquad \mbox{s.t. } \theta = \hat \theta^{(k+1)} \,. \nonumber
\end{align}
The following result provides the solution of the above problems.
\begin{proposition} \label{th:update_sigma_theta}
Introduce the matrix $D \in \mathbb{R}^{N^2\times N}$ such that $D a = \vect(\T_N(a))$, for any $a \in \mathbb{R}^{N}$.
Define
\begin{IEEEeqnarray}{rCl+rCl}
\hat A^{(k)} & = & D^T(R_1\hat S_{11}^{(k)}R_1^T \! \otimes I_N)D \, \\
\hat b^{(k)}  & = & \T_N(R_1 \hat s_{11}^{(k)})^T \tilde w_2 \,. \IEEEeqnarraynumspace
\end{IEEEeqnarray}
Then
\begin{equation} \label{eq:sol_theta}
\hat \theta^{(k+1)} = \argmin_{\theta} g_\theta^T \hat A^{(k)} g_\theta - 2 \hat b^{(k)T} g_\theta \,.
\end{equation}
The closed form updates of the noise variances are as follows
\begin{IEEEeqnarray}{rCll} \label{eq:sol_variances}
\hat \sigma^{2(k+1)}_1 & = &\frac{1}{N} & \left(\|\tilde w_1 - R_1 \hat s_{11}^{(k)}\|_2^2 + \tr{R_1 \hat P^{(k)} R_1^T}\right) \,,\nonumber\\
\hat \sigma^{2(k+1)}_2 & = &\frac{1}{N} & \left(\|\tilde w_2 - G_{\hat \theta^{(k+1)}} R_1 \hat s_{11}^{(k)}\|_2^2 \right.\nonumber \\
& & & \left. + \tr{G_{\hat \theta^{(k+1)}} R_ 1  \hat P^{(k)}R_1^T G_{\hat \theta^{(k+1)}}^T }\right) \,. \IEEEeqnarraynumspace
\end{IEEEeqnarray}
\end{proposition}
Each variance is the result of the sum of one term that measures the adherence of the identified systems to the data and one term that compensates for the bias in the estimates introduced by the Bayesian approach.
The update of the parameter $\theta$ involves a (generally) nonlinear least-squares problem, which can be solved using gradient descent strategies.
Note that, in case the impulse response $g_\theta$ is linearly parametrized
(e.g., it is an FIR system or orthonormal basis functions are used \citep{Wahlberg:90c}), then the update of $\theta$ is also available in closed-form.
\begin{example}
Assume that the linear parametrization $g_\theta = L\theta$, $L \in \mathbb{R}^{N\times n_\theta}$, is used, then
\begin{equation}
\hat \theta^{(k+1)} = \left(L^T \hat A^{(k)} L\right)^{-1} L^T \hat b^{(k)} \,.
\end{equation}
\end{example}
\subsection{Identification algorithm} \label{sec:algo}
The proposed method for module identification can be summarized in the following steps.
\begin{enumerate}
\item Find an initial estimate of $\hat \eta^{(0)}$, set $k = 0$.
\item Compute $\hat{s}_{11}^{(k)}$ and $\hat P^{(k)}$ from \eqref{eq:posterior_s}.
\item Update the kernel hyperparameters using \eqref{eq:thm_lambda}, \eqref{eq:thm_beta}.
\item Update the vector $\theta$ solving \eqref{eq:sol_theta}.
\item Update the noise variances from \eqref{eq:sol_variances}.
\item Check if the algorithm has converged. If not, set $k = k+1$ and go back to step 2.
\end{enumerate}

The method can be initialized in several ways.
One option is to first estimate $\hat{S}_{11}(q)$ by an empirical Bayes method using only $r_1$ and $\tilde w_1$.
Then, $\hat w_1$ is constructed from \eqref{eq:w_hat_first_stage}, using the obtained $\hat{S}_{11}(q)$.
Finally, $G$ is estimated using the prediction error method, using $\hat w_1$ as input and $\tilde w_2$ as output.

\subsection{Extension to general structures}
In this section, we generalize the previous algorithm to a general network structure with $m \le L$ reference signals $\{r_{l_1}(t),\,\ldots,\,r_{l_m}(t)\}$, and $p \le L$ modules $\{G_{j1}(q), \, \ldots,\, G_{jp}(q)\}$ sharing the same output $\tilde w_j(t)$ as the module of interest, and
modeled in time domain as $g_{\theta_1},\,\ldots,\,g_{\theta_p}$.
For any $i=1,\,\ldots,\,p$, we can write
\begin{equation} \label{eq:dyn_one_module}
\tilde w_{i} = R_{l_1} s_{il_1} + \ldots + R_{l_m} s_{il_m} + e_{k_i} = \Rbf s_i + e_{k_i} \,,
\end{equation}
where
$\Rbf \defeq [R_{l_1}\,\ldots\,R_{l_m}]$ and $s_i = [s_{il_1}^T\,\ldots\,s_{il_m}^T]^T$.
Using these definitions we can also write (cf. \eqref{eq:w_from_G})
\begin{equation} \label{eq:dyn_output}
\tilde w_j = r_j + G_{\theta_1}\Rbf s_1 + \ldots + G_{\theta_p}\Rbf s_p +e_j \,.
\end{equation}
Defining also $\wbf = [\tilde w_{1}^T \ldots \tilde w_{p}^T]^T$, $\sbf = [s_{1}^T\,\ldots\,s_{p}^T]^T$, $G_\theta = [G_{\theta_1}\,\ldots\,G_{\theta_p}]$, $\ebf_\wbf = [e_{1}^T \ldots e_{p}^T]^T$, we obtain the following expression for the network dynamics
\begin{align} \label{eq:network_1}
\wbf & = (I_p \otimes \Rbf) \sbf + \ebf_\wbf \nonumber\\
\tilde w_j - r_j & = G_{\theta}(I_p \otimes \Rbf) \sbf+e_j \,,
\end{align}
or, with $\zbf = [\wbf^T \, (\tilde w_j - r_j)^T]^T$
\begin{IEEEeqnarray}{rCl+rCl+rCl}
\zbf & = &
\Wbf_\theta \sbf + \ebf ,
&
\Wbf_\theta &=&
\begin{bmatrix}
(I_p \otimes \Rbf) \\
G_\theta (I_p \otimes \Rbf)
\end{bmatrix} ,
&
\ebf &=&
\begin{bmatrix}
\ebf_\wbf \\
e_j
\end{bmatrix} \,.
\IEEEeqnarraynumspace \label{eq:network}
\end{IEEEeqnarray}
Each sensitivity path $s_{il}$ is given a prior of the form \eqref{prior_s}, with hyperparameters $\lambda_{il}$ and $\beta_{il}$, assuming mutual independence between the sensitivity paths.
Although it may appear more sensible to incorporate some correlation among the sensitivity paths, at present, it is not clear how this can be done using Gaussian priors.
Some recent work suggests to enrich the stable spline kernel with a component
enforcing low McMillan degree \citep{prando2014bayesian}.
Furthermore, as we will see, assuming independent priors allows the kernel hyperparameters to be updated independently.
Introducing $\Lambda$ as the diagonal matrix with elements corresponding to $\{\lambda_{il}\}$, and similarly, defining $\Kbf_\beta$ with diagonal elements $\{K_{\beta_{il}} \}$, we have
\begin{equation}
p(\sbf;\,\Lambda,\,\Kbf_\beta) \sim \mathcal{N}\left(0,\,(\Lambda \otimes I_n) \Kbf_\beta\right)\,.
\end{equation}
We collect all the parameters characterizing the model into the vector $\eta$.
It follows that
\begin{equation}
p(\zbf;\,\eta) \sim \mathcal{N} (0,\,\Sigma_\zbf) \,,
\end{equation}
where $\Sigma_\zbf =  \Wbf_\theta (\Lambda \otimes I_n) \Kbf_\beta \Wbf_\theta^T + \Sigma_\ebf$, and
\begin{equation}
\Sigma_\ebf = \diag{\sigma_{1}^2,\,\ldots,\,\sigma_{p}^2,\,\sigma_{j}^2} \otimes I_N \,.
\end{equation}
Therefore, we can define the following ML criterion
\begin{equation} \label{eq:marg_lik_mimo}
\hat \eta = \argmax_{\eta} \log p(\zbf;\,\eta)\,.
\end{equation}

Having set the notation, we outline the ECM algorithm for this general setting below. To this end, note that
\begin{equation} \label{eq:post_s_mimo}
p(\sbf|\zbf;\,\eta) = \mathcal{N} (\hat \sbf,\, \Pbf) \,,
\end{equation}
where
\begin{align}
\hat \sbf & = \Pbf \Wbf_\theta^T \Sigma_\ebf^{-1} \zbf \,, \label{eq:post_mean_mimo} \\
\Pbf & = \left(\Wbf_\theta^T \Sigma_\ebf^{-1} \Wbf_\theta + \left((\Lambda \otimes I_n) \Kbf_\beta\right)^{-1}\right)^{-1}  \label{eq:post_cov_mimo} \,.
\end{align}
We use again the notation $\hat \sbf^{(k)},\,\hat \Pbf^{(k)},\,\hat \Sbf^{(k)}$ to mean the estimates of the corresponding quantities at iteration $k$.
\begin{proposition}
Let $\eta$ collect all the parameters characterizing \eqref{eq:marg_lik_mimo}, and let $\eta^{(k)}$ be its estimate after the $k$-th iteration of the ECM method. Then the estimate $\eta^{(k+1)}$ is obtained by means of the following updates.
\begin{enumerate}
\item[\emph{Hyperparameters:}] Define
\begin{equation} \label{eq:Q_beta_ji}
Q_{\beta_{ij}}(\beta) =  \log\det {K_\beta} + n\log \tr{K_{\beta}^{-1}\hat S_{ij}^{(k)}}\,,
\end{equation}
where $\hat S_{ij}^{(k)}$ is the $n \times n$ diagonal block of $\hat S^{(k)}$corresponding to the path $s_{ij}$. Then $\lambda_{ij}$  and $\beta_{ij}$ are updated as in Proposition \ref{th:update_hps}, for any $i,\,j$.
\item[\emph{Module parameters:}] Define
\begin{IEEEeqnarray}{rCl}
\hat \Abf^{(k)} & \coloneqq & (I_p \otimes D)^T\left( \bar \Rbf \hat \Sbf^{(k)} \bar \Rbf^T \otimes I_N\right)  (I_p \otimes D) \, , \IEEEeqnarraynumspace  \\
\hat \bbf^{(k)T} & \coloneqq & (\tilde w_j - r_j)^T\left[\T_N(\Rbf \hat s_1^{(k)}) \,\ldots\, \T_N(\Rbf \hat s_p^{(k)})\right] \,, \IEEEeqnarraynumspace
\end{IEEEeqnarray}
where $D$ is as in Proposition \ref{th:update_sigma_theta} and $\bar \Rbf = I_p \otimes \Rbf$.
Then
\begin{equation}
\hat \theta^{(k+1)} = \argmin_{\theta} \gbf_\theta^T \hat \Abf^{(k)} \gbf_\theta - 2\hat \bbf^{(k)T}\gbf_\theta \,,
\end{equation}
where $\gbf_\theta := [g_{\theta_1}^T \,\ldots \, g_{\theta_p}^T]^T$.
\item[\emph{Noise variances:}]
\begin{IEEEeqnarray}{rCl}
\hat \sigma^{2(k+1)}_{i} & = & \frac{1}{N} \left(\|\tilde w_{i} - \Rbf \hat s_i^{(k)}\|_2^2 + \tr{\Rbf \hat \Pbf_i^{(k)} \Rbf^T}\right) \IEEEeqnarraynumspace \nonumber\\
\hat \sigma^{2(k+1)}_j & = & \frac{1}{N} \left(\|\tilde w_j - r_j - G_{\hat \theta^{(k+1)}} (I\otimes \Rbf) \hat \sbf^{(k)}\|_2^2 \right.\nonumber \\
&  & \left. + \tr{G_{\hat \theta^{(k+1)}} \bar \Rbf  \hat \Pbf^{(k)}\bar \Rbf^T G_{\hat \theta^{(k+1)}}^T}\right) \,,
\end{IEEEeqnarray}
where $\Pbf_i^{(k)} $ is the $nm \times nm$ diagonal block of $\Pbf^{(k)}$, corresponding to the covariance matrix of $\hat s_i^{(k)}$.
\end{enumerate}
\end{proposition}

\section{Including additional sensors}
\label{sec:extension}
By using the kernel-based approach adopted above, the sensitivity paths could be modeled with only a few hyperparameters while still keeping the module of interest parametric.
One potential benefit with this approach is that including another reference signal will not increase the number of estimated parameters significantly.
Although the complexity of the problem increases slightly, only a few extra hyperparameters need to be estimated and the dimensions of \eqref{eq:sol_theta} remain the same in the update of $\theta$.

As reference signals can be added with little effort, a natural question is if also output measurements ``downstream'' of the module of interest can be added with little effort.
In Example~\ref{expl:network} the measurement $w_4$ is such a measurement that, with the same strategy as before, can be expressed as
\begin{equation}
w_4(t) = G_{43}(q) w_3(t) + r_4(t) \, .
\end{equation}
Using this measurement for the purpose of identification would require the identification of $G_{43}(q)$ in addition to the previously considered modules.
The signal $w_4(t)$ contains information about $w_3(t)$, and thus information about the module of interest. The price we have to pay for this information is the additional parameters to estimate and, as we will see, another layer of complexity.

\begin{figure}
    \centering
  \begin{tikzpicture}[scale=1,>=triangle 60,tip/.style={->,shorten >=-0.5pt},every join/.style={rounded corners,shorten >=-0.6pt}] 
\matrix[row sep={6mm,between origins},column sep={12mm,between origins}] {
\node (p11) [point,label={$r_1$}] {};
  & \node (S11) [block,minimum width=3.2em] {$S_{11}(q)$};
  & \node (w1) [sum,label={[yshift=-1mm,xshift=3mm]$w_1$}] {$+$};
  & \node (G) [block,minimum width=3.2em] {$G_{21}(q)$};
  & \node (w2) [sum,label={[yshift=-1mm,xshift=3mm]$w_2$}] {$+$};
  & \node (F) [block,minimum width=3.2em] {$F(q)$};
  & \node (w3) [sum,label={[yshift=-1mm,xshift=3mm]$w_3$}] {$+$}; \\
};

{ [start chain]
  \chainin (p11);
  \chainin (S11)  [join=by tip];
  \chainin (w1)   [join=by tip];
  \chainin (G)    [join=by tip];
  \chainin (w2)   [join=by tip];
  \chainin (F)    [join=by tip];
  \chainin (w3)   [join=by tip];
}
\end{tikzpicture}
  \caption{Basic network of 1 reference signal and 3 internal variables.}
   \label{fig:nebx-network}
\end{figure}
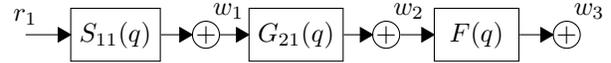

To extend the previous framework to include additional measurements after the module of interest, let us consider the case where we would like to include only one additional measurement, in this context denoted by $\tilde w_3(t)$; the generalization to more sensors is straightforward but notationally heavy. Let the path linking the target module to the additional sensor be denoted by $F(q)$, with impulse response $f$.
Furthermore, let us for simplicity consider the one-reference-signal-one-input case
again, \ie \eqref{eq:dynamcs_2}, \eqref{eq:dynamcs_3}.
The setting we consider has been illustrated in Figure~\ref{fig:nebx-network}.
We model also this module using a Bayesian framework by interpreting $f$ as a zero-mean Gaussian stochastic vector, i.e.,
\begin{equation}
p(f; \lambda_f, K_{\beta_f}) \sim \mathcal{N} (0, \lambda_f K_{\beta_f}) \, ,
\end{equation}
where again $K_{\beta_f}$ is the first-order stable spline kernel \eqref{eq:stablespline}.
We introduce the following variables
\begin{IEEEeqnarray}{rCl}
\sigma &=& \begin{bmatrix}
\sigma_{1}^2  & \sigma_{2}^2 & \sigma_{3}^2
\end{bmatrix} \, ,\\
z &=& \begin{bmatrix}
\tilde w_1^T & \tilde w_2^T & \tilde w_3^T
\end{bmatrix}^T \, ,\\
z_f &=& \tilde w_3 \, .
\end{IEEEeqnarray}
For given vales of $\theta$, $s$ and $f$, we construct
\begin{IEEEeqnarray}{rCl}
W_s &=& \begin{bmatrix}
R \\ G_{\theta}R \\ F G_\theta R
\end{bmatrix} \, , \label{eq:W_theta_additional} \\
W_f &=& \T_N (G_\theta Rs_{11}) \, \label{eq:W_f_additional},
\\
\Sigma &=& \diag{\sigma} \kronecker I_N \, .
\end{IEEEeqnarray}
Notice that the last
internal variable $w_3$ can be expressed as
\begin{IEEEeqnarray*}{rCl}
w_3 & = & F G_\theta S_{11} r \\
&=& G_\theta F S_{11} r \\
&=& G_\theta F R s_{11} \\
&=& G_\theta R F s_{11} \\
&=& G_\theta R v, \yesnumber \label{eq:v-def}
\end{IEEEeqnarray*}
where commutation of the matrices follows from the fact that they are lower-triangular Toeplitz matrices, and $v := Fs_{11}$. For ease of exposition, we will also use the notation $v = f \ast s_{11}$.

The key difficulty in this setup is that the description of the measurements and the system
description with both $s_{11}$ and $f$ no longer admit a jointly Gaussian
probabilistic model, because $v$ in \eqref{eq:v-def} is the result of the convolution of two Gaussian vectors. 
In fact, a closed-form expression is not available. This fact has a detrimental effect in our empirical Bayes approach, because the marginal likelihood estimator of
$$
\eta = \begin{bmatrix} \sigma & \lambda_s & \beta_s & \lambda_f & \beta_f & \theta \end{bmatrix},
$$
where $\lambda_s,\,\beta_s$ are the hyperparameters of the prior of $s_{11}$, that is
\begin{IEEEeqnarray}{rCl}
\hat \eta &=& \argmax_\eta p(z; \eta) \\
&=& \argmax_\eta \int p(z, s_{11}, f; \eta)\, \text{d}s_{11} \, \text{d}f ,
\label{eq:EM-integral}
\end{IEEEeqnarray}
does not admit an analytical expression, since the integral \eqref{eq:EM-integral} is intractable. To treat this problem, again we resort to EM-type methods. In this case, the latent variables to add to the problem are both $s_{11}$ and $f$, so that the EM method has to alternate between the following steps.
\begin{enumerate}
\item[(E-step)] Given an estimate $\hat \eta^{(k)}$ (computed at the $k$-th iteration of the algorithm), compute
    \begin{equation} \label{eq:nebx_estep}
    Q^{(k)}(\eta) \defeq \mathbb{E} \left[\log p(z,\,s_{11},\,f;\,\eta) \right] \,,
    \end{equation}
    where the expectation is taken with respect to the target distribution when the estimate $\eta^{(k)}$ is used, i.e., $p(s_{11}, f|z,\,\hat \eta^{(k)})$\,;
\item[(M-step)] Solve the problem
    \begin{equation} \label{eq:nebx_Q_function}
    \hat \eta^{(k+1)} = \argmax_{\eta}  Q^{(k)}(\eta) \,.
    \end{equation}
\end{enumerate}
While the M-Step remains substantially unchanged, the E-step requires more attention. Now, it requires the computation of the integral
\begin{align}
\mathbb{E} & \left[\log p(z,\,s_{11},\,f;\,\eta) \right]  =  \\
& \int \log p(z,\,s_{11},\,f;\,\eta) p(s_{11}, f|z,\,\hat \eta^{(k)}) \, \text{d}s_{11} \, \text{d}f , \nonumber
\end{align}
which does not admit an analytical solution, because the posterior distribution $p(s_{11}, f|z,\,\hat \eta^{(k)})$ is non-Gaussian (it does not have an analytical form, in fact).
However, using Markov Chain Monte Carlo (MCMC) techniques, we can compute an approximation of the integral by sampling from the joint posterior density (also called a target distribution)
\begin{equation}
p(s_{11},f|z; \eta).
\label{eq:target-distribution}
\end{equation}
As pointed out before, \eqref{eq:target-distribution} does not admit a closed-form expression and hence direct sampling is a hard task.
However, if it is easy to draw samples from the conditional probability distributions, samples of \eqref{eq:target-distribution} can be easily drawn using the Gibbs sampler.
In Gibbs sampling, each conditional is considered the state of a Markov chain; by iteratively drawing samples from the conditionals, the Markov chain will converge to its stationary distribution, which corresponds to the target distribution.
In our problem, the conditionals of \eqref{eq:target-distribution} are as follows
\begin{itemize}
\item $p(s_{11}|f,z;\eta)$. Using \eqref{eq:W_theta_additional}, we write the linear model
\begin{equation}
z = W_s s_{11} + e,
\end{equation}
where $e = [e_1^T\,e_2^T\,e_3^T]^T$. Then, given $f$, the vectors $s_{11}$ and $z$ are jointly Gaussian, so that
\begin{equation}
p(s|f,z;\eta) \sim \mathcal{N} (m_s, P_s) \, ,
\end{equation}
with
\begin{IEEEeqnarray*}{rCl}
P_s &=& \left(  W_s^T \Sigma^{-1}W_s + (\lambda_s K_{\beta_{s}})^{-1} \right)^{-1} \\
m_s &=& P_s W_s^T \Sigma^{-1}z \, .
\end{IEEEeqnarray*}
\item $p(f|s,z;\eta)$.
Given $s$ and $r$, all sensors but the last becomes redundant. Using \eqref{eq:W_f_additional} we write the linear model
\begin{equation}
z_f = W_f f + e_3,
\end{equation}
which shows that
\begin{equation}
p(f|s_{11},z;\eta) \sim \mathcal{N} (m_f, P_f),
\end{equation}
with
\begin{IEEEeqnarray*}{rCl}
P_f &=& \left(  \frac{W_f^T W_f}{\sigma_3^2} + (\lambda_f K_{\beta_f})^{-1} \right)^{-1} \\
m_f &=& P_f \frac{W_f^T}{\sigma_3^2} z_f \, .
\end{IEEEeqnarray*}
\end{itemize}
The following algorithm summarizes the Gibbs sampler used for dynamic network identification.
\begin{alg} \label{alg:gibbs}
Gibbs sampler for a dynamic network.
Initialization: compute initial value of $s^{0}$ and $f^{0}$.
For $k=1$ to $M+M_0$:
\begin{enumerate}

  \item Draw the sample $s^{k}$ from $p(s|f^{k-1},z;\eta)$;

  \item Draw the sample $f^{k}$ from $p(f|s^{k},z;\eta)$;

\end{enumerate}

\end{alg}
In this algorithm, $M_0$ is the number of initial samples that are discarded, which is also known
as the \emph{burn-in} \citep{Meyn2009}. These samples are discarded since the
Markov chain needs a certain number of samples to converge to its stationary distribution.

\subsection{The ECM method with additional sensor}

We now discuss the computation of the E-step and the CM-steps using the Gibbs sampler scheme introduce above.

\begin{proposition} \label{thm:E-step-gibbs}
Introduce the mean and covariance quantities
\begin{IEEEeqnarray}{rCl}
s_s^M &=& \frac{1}{M} \sum_{k=M_0+1}^{M_0+M} s^k \,,\label{eq:gibbs1}\\
f_s^M &=& \frac{1}{M} \sum_{k=M_0+1}^{M_0+M} f^k \,, \\
v_s^M &=& \frac{1}{M} \sum_{k=M_0+1}^{M_0+M} v^k \,, \\
P_s^M &=& \frac{1}{M} \sum_{k=M_0+1}^{M_0+M} (s^k-s^M_s)(s^k-s^M_s)^T \,,\\
P_f^M &=& \frac{1}{M} \sum_{k=M_0+1}^{M_0+M} (f^k-f^M_s)(f^k-f^M_s)^T \,, \\
P_v^M &=& \frac{1}{M} \sum_{k=M_0+1}^{M_0+M} (v^k-v^M_s)(v^k-v^M_s)^T \,,
\label{eq:gibbs2}
\end{IEEEeqnarray}
where $s^k$, $f^k$ and $v^k = s^k \ast f^k $ are samples drawn using Algorithm~\ref{alg:gibbs}.

Define
\begin{IEEEeqnarray*}{rCl}
\tilde Q_s(\lambda, \beta, x, X)
	& \defeq & \log \det \lambda K_\beta \\
    &   & + \tr{ (\lambda K_\beta)^{-1}(xx^T + X)} \, , \\
\tilde Q_z(\sigma^2, z, x, X) & \defeq &  N \log \sigma^2 + \frac{1}{\sigma^2}
\norm{z-Rx}_2^2  \\
&& + \frac{1}{\sigma^2} \tr{ R XR^T } \, ,
\\
\tilde Q_f(\sigma^2, z, \theta, x, X) & \defeq &  N \log \sigma^2 + \frac{1}{\sigma^2}
\norm{z-G_\theta Rx}_2^2  \\
&& + \frac{1}{\sigma^2} \tr{ G_\theta R XR^T G_\theta^T } \, .
\end{IEEEeqnarray*}
Then
\begin{IEEEeqnarray*}{rCl}
-2Q^{(k)}(\eta) &=& \lim_{M \to \infty} \tilde Q_s(\lambda_s, \beta_s, s_s^M,
P_s^M) \, ,\\
&&+ \tilde Q_s(\lambda_f, \beta_f, f_s^M, P_f^M)             \, , \\
&&+ \tilde Q_z(\sigma_1^2, \tilde w_1, s_s^M, P_s^M)         \, , \\
&&+ \tilde Q_f(\sigma_2^2, \tilde w_2, \theta, s_s^M, P_s^M) \, ,\\
&&+ \tilde Q_f(\sigma_3^2, \tilde w_3, \theta, v_s^M, P_v^M) \, .
\yesnumber \label{eq:Q-gibbs}
\end{IEEEeqnarray*}
\end{proposition}

The CM-steps are now very similar to the previous method and follows by similar
reasoning as in the proof of Proposition~\ref{th:update_sigma_theta}.

\begin{proposition}
\label{prop:M-step}
Let $\hat \eta^{(k)}$ be the parameter estimate obtained at the $k$:th iteration.
Define $ S^M_s = s^{M}_s ( s^{M}_s)^T +  P^M_s$,
$ S^{M}_v = v^{M}_s ( v^{M}_s)^T + P^M_v$,
\begin{IEEEeqnarray*}{rCl}
\hat A_s & = &  D^T(R S_{s}^{M}R^T \! \otimes I_N)D \,, \\
\hat A_v & = &  D^T(R S_{v}^{M}R^T \! \otimes I_N)D \,, \\
\hat b_s & = &  \T_N(R s_s^{M})^T \tilde w_2      \,, \\
\hat b_v & = &  \T_N(R v_s^{M})^T \tilde w_3     \,. \IEEEeqnarraynumspace
\end{IEEEeqnarray*}
Then the updated parameter vector $\hat \eta^{(k+1)}$ is obtained as follows
\begin{IEEEeqnarray}{rCl} \label{eq:nebx-sol_theta}
\hat \theta^{(k+1)} &=& \argmin_{\theta}
g_\theta^T\left(\frac{1}{\sigma_2^2} \hat A_s  + \frac{1}{\sigma_3^2} \hat A_v\right) g_\theta
\nonumber \\
&& -2 \left(\frac{1}{\sigma_2^2} \hat b_s^T
 + \frac{1}{\sigma_3^2} \hat b_v^T\right) g_\theta \,.
\end{IEEEeqnarray}
The closed form updates of the noise variances are
\begin{IEEEeqnarray}{rCl} \label{eq:nebx-sol_variances}
  \hat \sigma^{2(k+1)}_1
    & = & \frac{1}{N} \left(\|\tilde w_1 - Rs_s^M\|_2^2 + \tr{R P_s^M R^T}
      \right) \,,\nonumber \\
  \hat \sigma^{2(k+1)}_2
    & = &\frac{1}{N} \left( \vphantom{s_s^M G_{\hat \theta^{(k+1)}}^T} \|
      \tilde{w}_2 - G_{\hat \theta^{(k+1)}} R s_s^M \|_2^2 \right.\nonumber \\
    &   &  \phantom{\frac{1}{N}} + \left. \tr{G_{\hat \theta^{(k+1)}} R
      P_s^MR^T G_{\hat \theta^{(k+1)}}^T } \vphantom{s_s^M
      G_{\hat \theta^{(k+1)}}^T} \right) \, , \nonumber \\
  \hat \sigma^{2(k+1)}_3
    & = & \frac{1}{N} \left(\|\tilde w_3 - G_{\hat \theta^{(k+1)}} R v_s^M \|_2^2
      \vphantom{s_s^M G_{\hat \theta^{(k+1)}}^T} \right.\nonumber \\
    &   & \phantom{\frac{1}{N}} + \left. \tr{G_{\hat \theta^{(k+1)}} R  P_v^M R^T
      G_{\hat \theta^{(k+1)}}^T } \vphantom{s_s^M G_{\hat \theta^{(k+1)}}^T}
      \right) \,. \IEEEeqnarraynumspace
\end{IEEEeqnarray}
The kernel hyperparameters are updated through \eqref{eq:thm_beta} and \eqref{eq:thm_lambda} for both $s_{11}$ and $f$.
\end{proposition}

\subsection{Identification algorithm} \label{sec:algo_nebx}
The proposed method for module identification can be summarized in the following steps.
\begin{enumerate}
\item Find an initial estimate of $\hat \eta^{(0)}$, set $k = 0$.
\item Compute the quantities \eqref{eq:gibbs1}-\eqref{eq:gibbs2} using Algorithm 1.
\item Update the kernel hyperparameters using \eqref{eq:thm_lambda}, \eqref{eq:thm_beta}.
\item Update the vector $\theta$ solving \eqref{eq:nebx-sol_theta}.
\item Update the noise variances from \eqref{eq:nebx-sol_variances}.
\item Check if the algorithm has converged. If not, set $k = k+1$ and go back to step 2.
\end{enumerate}
As can be seen, the main difference with the one-input-one-sensor algorithm (see Section \ref{sec:algo_nebx}) is that Step 2 of the algorithm requires a heavier computational burden, because of the integration via Gibbs sampling. Nevertheless, as will be seen in the next section, this pays off in terms of performance in identifying the target module.

\section{Numerical experiments} \label{sec:experiments}

In this section, we present results from two Monte Carlo simulations to
illustrate the performance of the proposed method, which we abbreviate as
\emph{Network Empirical Bayes} (NEB) and its extension NEBX outlined in
Section~\ref{sec:extension}, and we compare with SMPE (see Section \ref{sec:SMPE}).
We consider the network case of Example~\ref{expl:network} and a simple closed loop network.
The reference signals used are zero-mean unit-variance Gaussian white noise.
The noise signals $e_k$ are zero-mean Gaussian white noise with variances such
that noise to signal ratios $\var w_k/\var e_k$ are constant.
The setting of the compared methods
are provided in some more details below, where the model order of the plant
$G(q)$ is known for both the SMPE method and the proposed NEB method.

\emph{NEB:}
The method is initialized by the two-stage method.
First, $\hat{S}(q)$ is estimated by least-squares.
Second, $G$ is estimated using MORSM \citep{MORSM} from the simulated
signal $\hat w$ obtained from \eqref{eq:w_hat_first_stage} and $\tilde w_j$.
MORSM is an iterative method that is asymptotically efficient for open loop data.
Then, the iterative method outlined in Section \ref{sec:algo}
is employed with the stopping criteria
$\norm{\hat \eta^{(k+1)} - \hat \eta^{(k)}}/\norm{\hat \eta^{(k)}} < 10^{-10}$.

\emph{NEBX:}
The method is initialized by NEB. $f^0$ is obtained by an empirical Bayes method
using simulated input and measured output of $f$.
Then, the iterative method outlined in Section \ref{sec:extension}
is employed with the stopping criteria
$\norm{\hat \eta^{(k+1)} - \hat \eta^{(k)}}/\norm{\hat \eta^{(k)}} < 10^{-10}$,
or a maximum of $50$ iterations.

\emph{SMPE:}
The method is initialized by the two-stage method, exactly as NEB.
Then, the cost function \eqref{eq:V_SMPE}, with a slight modification, is minimized.
The modification of the cost function comes from that, as mentioned before, the
SMPE method assumes that the noise variances are known.
To make the comparison fair, also the noise variances need to be estimated.
By maximum likelihood arguments, the logarithm of the
determinant of the complete noise covariance matrix is added to the cost
function \eqref{eq:V_SMPE} and the noise variances are included in $\theta$,
the vector of parameters to estimate. The tolerance is set to
$\norm{\hat \theta^{(k+1)} - \hat \theta^{(k)}}/\norm{\hat \theta^{(k)}} < 10^{-10}$.

The simulations were run in Julia, a high-level, high-performance dynamic programming language for technical computing \citep{Bezanson_2017}.

\subsection{Closed-loop identification}

The first Monte Carlo simulation is from a system operating in closed loop with
an unknown low order controller with $N=200$ data samples. This setting is
slightly different to the standard closed-loop setting in that
the measurement noise of $\tilde w_2$ is not fed back in the loop, and that the
signals $w_1$ and $w_2$ are treated completely symmetric.
The noise to signal ratio are all set to $1$. The true plant and
true controller are chosen such that the sensitivity function $S(q^{-1})$ has an
impulse response that can be well approximated by $n=100$ impulse response coefficients.
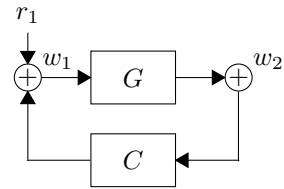
\begin{figure}[!ht]
    \centering
  \begin{tikzpicture}[scale=1,>=triangle 60,tip/.style={->,shorten >=-0.5pt},every join/.style={rounded corners,shorten >=-0.6pt}] 
\matrix[row sep={6mm,between origins},column sep={5mm,between origins}] {
\node (p11) [point,label={[yshift=-0.0mm,xshift=0mm]$r_1$}] {};		& [9mm]	& [9mm]
\\    
\node (w1) [sum,label={[yshift=-2mm,xshift=4mm]$w_1$}] {$+$}; & 
\node (G) [block,minimum width=3.2em] {$G$}; &
\node (w2) [sum,label={[yshift=-2mm,xshift=4mm]$w_2$}] {$+$};
\\
[5mm]
\node (p31) [point] {}; &
\node (C) [block,minimum width=3.2em] {$C$}; &
\node (p33) [point] {};
\\
};

{ [start chain]
	\chainin (p11);
	\chainin (w1)		[join=by tip];
	\chainin (G) 		[join=by tip];
	\chainin (w2) 		[join=by tip];
	\chainin (p33)		[join];
	\chainin (C)		[join=by tip];
	\chainin (p31)		[join];
	\chainin (w1)		[join=by tip];
}

\end{tikzpicture}
  \caption{Closed loop network of first Monte Carlo simulation.}
   \label{fig:closed_loop}
\end{figure}
The closed loop is depicted in Figure~\ref{fig:closed_loop}, where
\begin{IEEEeqnarray*}{rCl+rCl}
G(q,\theta) &=& \frac{b_1 q^{-1} + b_2 q^{-2}}{1 + a_1 q^{-1} + a_2 q^{-2} },
\IEEEeqnarraynumspace  \yesnumber 
\end{IEEEeqnarray*}
The controller $C$ is given by
\begin{IEEEeqnarray*}{rCl}
C(q,\theta) &=& \frac{0.8 + 0.4 q^{-1} - 0.5 q^{-2}}{1 + 0.5 q^{-1} + 0.2 q^{-2} },
\IEEEeqnarraynumspace  \yesnumber 
\end{IEEEeqnarray*}
with the parameter vector {$\theta = [b_1, \, b_2, \, a_1, \, a_2]$}, and true
parameters $ \theta^0 = [0.4, \, 0.5, \, -0.4, \, 0.3]$.

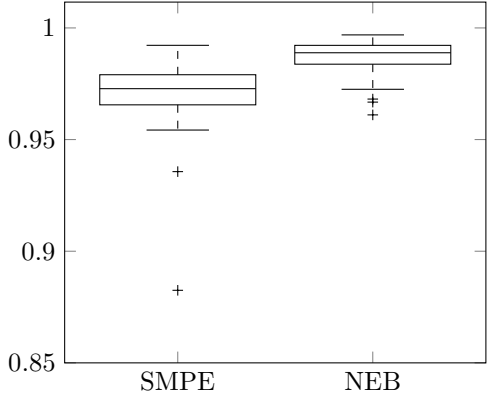
\begin{figure}[!ht]
    \centering
  \begin{tikzpicture}
	\begin{axis}[
			width=0.8\figurewidth,
			height=0.8\figureheight,
			at={(0\figurewidth,0\figureheight)},
			scale only axis,
			boxplot/draw direction=y,
			xtick={1,2,3},
			ymin = 0.85,
			xticklabels={SMPE, NEB},]
		\addplot[boxplot,mark=+] table[y index=0]{smpecl.csv};
		\addplot[boxplot,mark=+] table[y index=0]{nebcl.csv};
	\end{axis}
\end{tikzpicture}
  \caption{Box plot of the fit of the impulse response of $G$
          obtained by the SMPE, and NEB methods respectively.}
   \label{fig:boxplot-G}
\end{figure}

\begin{figure}[!ht]
    \centering
%

\begin{tikzpicture}

  \begin{axis}[%
  width=0.8\figurewidth,
  height=0.8\figureheight,
  at={(0\figurewidth,0\figureheight)},
  scale only axis,
  xmin=0.84,
  xmax=1,
  xtick={0.85, 0.9, 0.95, 1},
  xlabel={$\text{FIT}_{\text{SMPE}}$},
  ymin=0.94,
  ymax=1,
  ytick={0.95, 1},
  ylabel={$\text{FIT}_{\text{NEB}}$},
  axis background/.style={fill=white},
  legend style={at={(0.05,0.05)},
  		anchor=south west}
  ]
    \addplot[color=blue,only marks,mark=asterisk] %
      table [col sep=comma]{twovsnebcl.csv};
    \addplot [color=black,solid]
      table[row sep=crcr]{%
    0.1	0.1\\
    0.15	0.15\\
    0.2	0.2\\
    0.25	0.25\\
    0.3	0.3\\
    0.35	0.35\\
    0.4	0.4\\
    0.45	0.45\\
    0.5	0.5\\
    0.55	0.55\\
    0.6	0.6\\
    0.65	0.65\\
    0.7	0.7\\
    0.75	0.75\\
    0.8	0.8\\
    0.85	0.85\\
    0.9	0.9\\
    0.95	0.95\\
    1	1\\
    };
    \legend{NEB}
  \end{axis}
\end{tikzpicture}%
  \caption{Each fit of the impulse response coefficients of $G$ for NEB
  compared with SMPE for 100 Monte Carlo simulations. The black line
  represents $y = x$, \ie when SMPE performs equally good as NEB.
  Note the scaling of the x-axis of this figure.}
  \label{fig:fit-smpe-vs-neb}
\end{figure}
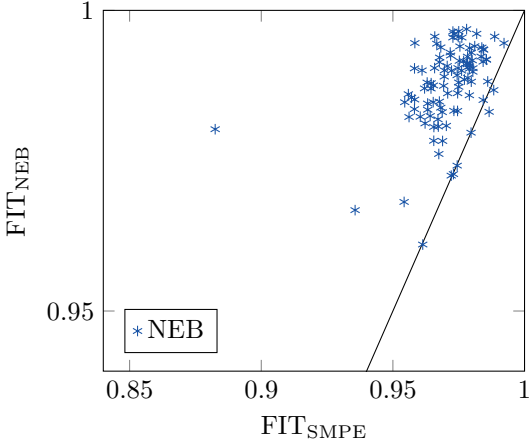

The two methods are compared using the fit of the impulse response coefficients
of $g$ according to
\begin{IEEEeqnarray*}{rCl}
FIT &=& 1- \frac{\norm{g^0 - \hat g }_2}{\norm{g^0}_2}
\IEEEeqnarraynumspace  \yesnumber \label{eq:fit}
\end{IEEEeqnarray*}
For this example, the proposed NEB method achieves a higher fit, on average,
than the SMPE
method, cf. the box plot of Figure~\ref{fig:boxplot-G}. Comparing the fits obtained
at each Monte Carlo run (see Figure~\ref{fig:fit-smpe-vs-neb}),
it can be seen that NEB consistently performs at least as good as
SMPE for almost every Monte Carlo run and in some runs considerably better.
From the sample means and variance reported in Table~\ref{tbl:thetaG},
it can be seen that, in general, the estimates produced by NEB have
smaller variance than SMPE while their mean values are similar.

\begin{table*}
  \caption{Sample mean and sample variance of the parameters estimates for
  $\hat G$ for compared methods.}
  \label{tbl:thetaG}
  \begin{center}
    \begin{tabular}{ l | *{2}{c} | *{2}{c} | *{2}{c} | *{2}{c} }
      \multicolumn{1}{ c |}{} & \multicolumn{2}{ c |}{$b_1^0 = 0.2$}
      & \multicolumn{2}{ c |}{$b_2^0 = 0.3$}  & \multicolumn{2}{ c |}{$a_1^0 = 0.4$}
      & \multicolumn{2}{ c }{$a_2^0 = 0.5$} \\
      Method & $\mathrm{E}_s \, \hat b_1$ & $N \cdot \mathrm{Var}_s \, \hat b_1 $
      & $\mathrm{E}_s \, \hat b_2$ & $N \cdot \mathrm{Var}_s \, \hat b_2$
      & $\mathrm{E}_s \, \hat a_1$ & $N \cdot \mathrm{Var}_s \, \hat a_1$
      & $\mathrm{E}_s \, \hat a_2$ & $N \cdot \mathrm{Var}_s \, \hat a_2$ \\
      \hline
      SMPE & 0.21 & 0.43 & 0.31 & 0.93 & 0.50 & 3.4 & 0.16 & 2.8
      \\
      NEB  & 0.20 & 0.22 & 0.31 & 0.26 & 0.68 & 2.9 & 0.23 & 2.0
      \\ \hline
    \end{tabular}
  \end{center}
\end{table*}


\subsection{Dynamic network example}
This Monte Carlo simulation compares the NEB method and NEBX with the SMPE method
on data from the network of Example~\ref{expl:network},
illustrated in Figure~\ref{fig:network}, where each of
the modules are of second order, \ie
\begin{equation*}
G_{ij}(q) = \frac{ b_1q^{-1} + b_2q^{-2} }{ 1 + a_1q^{-1} + a_2q^{-2}},
\end{equation*}
for a set of parameters that were chosen such that all modules are
stable and $\{S_{12}(q),S_{24}(q),S_{22}(q),S_{24}(q)\}$ are stable
and can be well approximated with 70 impulse response coefficients.
Two reference signals, $r_2(t)$ and $r_4(t)$ are available and $N=200$
data samples are used with the goal to estimate $G_{31}(q)$ and $G_{32}$.
In total 6 transfer functions are estimated,
$\{S_{12}(q),S_{24}(q),S_{22}(q),S_{24}(q), G_{31}(q)$ and
$G_{32}(q)\}$, where $\{S_{12}(q),S_{24}(q),S_{22}(q),S_{24}(q)\}$ are
each parameterized by $n=75$ impulse response coefficients in all methods.
For NEBX also $G_{43}(q)$ is estimated by $n=75$ impulse response coefficients.
The noise to signal ratio at each measurement is set to $\var w_k/\var e_k = 0.1$
and the additional measurement used in NEBX has a lower noise to signal ratio
of $\var w_4/\var e_4 = 0.01$.

\begin{table*}
  \caption{Sample mean and sample variance of the parameters estimates for
  $\hat G_{31}$ for the three compared methods.}
  \label{tbl:thetaG31}
  \begin{center}
    \begin{tabular}{ l | *{2}{c} | *{2}{c} | *{2}{c} | *{2}{c}}
      \multicolumn{1}{ c |}{} & \multicolumn{2}{ c |}{$b_1^0 = 0.2$}
      & \multicolumn{2}{ c |}{$b_2^0 = 0.3$}  & \multicolumn{2}{ c |}{$a_1^0 = 0.4$}
      & \multicolumn{2}{ c }{$a_2^0 = 0.5$} \\
      Method & $\mathrm{E}_s \, \hat b_1$ & $N \cdot \mathrm{Var}_s \, \hat b_1 $
      & $\mathrm{E}_s \, \hat b_2$ & $N \cdot \mathrm{Var}_s \, \hat b_2$
      & $\mathrm{E}_s \, \hat a_1$ & $N \cdot \mathrm{Var}_s \, \hat a_1$
      & $\mathrm{E}_s \, \hat a_2$ & $N \cdot \mathrm{Var}_s \, \hat a_2$ \\
      \hline
      SMPE & 0.20 & 0.088 & 0.28 & 0.075 & 0.36 & 1.6 & 0.53 & 0.85
      \\
      NEB  & 0.21 & 0.049 & 0.29 & 0.070 & 0.36 & 0.94 & 0.52 & 0.62
      \\
      NEBX & 0.20 & 0.024 & 0.29 & 0.036 & 0.40 & 0.60 & 0.50 & 0.52
      \\ \hline
    \end{tabular}
  \end{center}
\end{table*}

\begin{table*}
  \caption{Sample mean and sample variance of the parameters estimates for
  $\hat G_{32}$ for the three compared methods.}
  \label{tbl:thetaG32}
  \begin{center}
    \begin{tabular}{ l | *{2}{c} | *{2}{c} | *{2}{c} | *{2}{c} }
      \multicolumn{1}{ c |}{} & \multicolumn{2}{ c |}{$b_1^0 = 0.4$}
      & \multicolumn{2}{ c |}{$b_2^0 = 0.5$}  & \multicolumn{2}{ c |}{$a_1^0 = 0.5$}
      & \multicolumn{2}{ c }{$a_2^0 = 0.15$} \\
      Method & $\mathrm{E}_s \, \hat b_1$ & $N \cdot \mathrm{Var}_s \, \hat b_1 $
      & $\mathrm{E}_s \, \hat b_2$ & $N \cdot \mathrm{Var}_s \, \hat b_2$
      & $\mathrm{E}_s \, \hat a_1$ & $N \cdot \mathrm{Var}_s \, \hat a_1$
      & $\mathrm{E}_s \, \hat a_2$ & $N \cdot \mathrm{Var}_s \, \hat a_2$ \\
      \hline
      SMPE & 0.34 & 1.9 & 0.44 & 2.1 & 0.60 & 5.0 & 0.23 & 3.0
      \\
      NEB  & 0.34 & 0.30 & 0.44 & 0.30 & 0.65 & 1.0 & 0.26 & 0.84
      \\
      NEBX & 0.36 & 0.11 & 0.45 & 0.16 & 0.63 & 0.68 & 0.25 & 0.55
      \\ \hline
    \end{tabular}
  \end{center}
\end{table*}

The fits of the impulse responses of $G_{31}$ and $G_{32}$ for the experiment
are shown as a boxplot in Figure~\ref{fig:boxplot-G31} and
Figure~\ref{fig:boxplot-G32} respectively.
Comparing the fits obtained
at each Monte Carlo run (see Figure~\ref{fig:fit-smpe-vs-neb2} and
Figure~\ref{fig:fit-smpe-vs-neb2}),
the proposed NEB and NEBX methods are competitive with the SMPE method for this network.
In many cases, the SMPE method failed to produce a reasonable estimate as 10 percent of the
Monte Carlo runs gave a negative fit and were removed before the impulse response
fits, boxplots
and parameter sample means and variances were computed. From the sample means and
variance reported in Table~\ref{tbl:thetaG31} and Table~\ref{tbl:thetaG32},
it can be seen that, in general, the estimates produced by NEB and NEBX have,
in general, significantly smaller variance than SMPE, while the mean values are
roughly the same. Recalling that one of the motivations of the proposed methods
was to reduced the variance induced by the high order modeling of the sensitivity paths,
both the closed-loop example and network example gives some support for this motivation.

In almost all of the Monte Carlo runs, NEBX
outperformed NEB in this simulation.
However, NEBX is significantly more computationally expensive than NEB.

\begin{figure}[!ht]
    \centering
  \begin{tikzpicture}
	\begin{axis}[
			width=0.8\figurewidth,
			height=0.8\figureheight,
			at={(0\figurewidth,0\figureheight)},
			scale only axis,
			boxplot/draw direction=y,
			xtick={1,2,3},
			ymin = 0.8,
			xticklabels={SMPE, NEB, NEBX},]
		\addplot[boxplot,mark=+] table[y index=0]{smpenl1.csv};
		\addplot[boxplot,mark=+] table[y index=0]{nebnl1.csv};
		\addplot[boxplot,mark=+] table[y index=0]{nebxnl1.csv};
	\end{axis}
\end{tikzpicture}
  \caption{Box plot of the fit of the impulse response of $G_{31}$
          obtained by the methods SMPE, NEB and NEBX respectively.}
   \label{fig:boxplot-G31}
\end{figure}
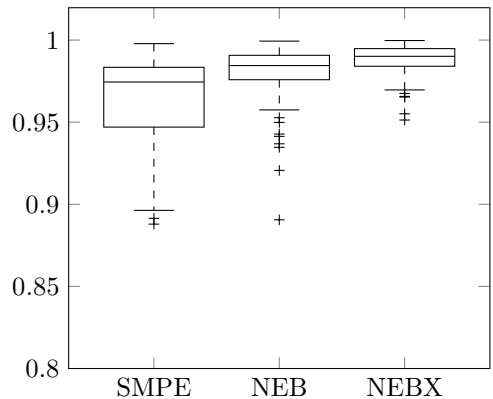
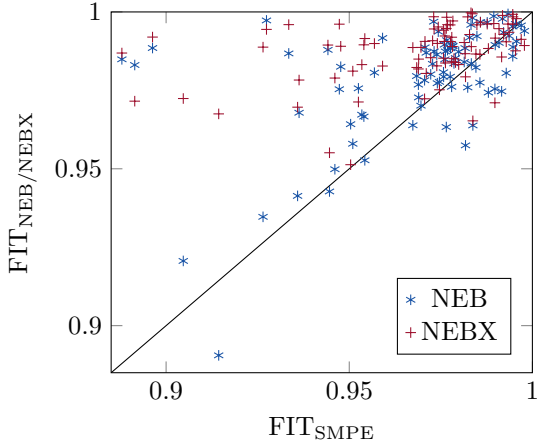
\begin{figure}[!ht]
    \centering
%

\begin{tikzpicture}

\begin{axis}[%
width=0.8\figurewidth,
height=0.8\figureheight,
at={(0\figurewidth,0\figureheight)},
scale only axis,
xmin=0.885,
xmax=1,
xtick={0.9, 0.95, 1},
xlabel={$\text{FIT}_{\text{SMPE}}$},
ymin=0.885,
ymax=1,
ytick={0.9, 0.95, 1},
ylabel={$\text{FIT}_{\text{NEB/NEBX}}$},
axis background/.style={fill=white},
legend style={at={(0.95,0.05)},
		anchor=south east}
]
\addplot[only marks,color=blue,mark=asterisk,mark size=2] %
  table [col sep=comma]{twovsneb.csv};
\addplot[only marks,color=red,mark=+,mark size=2] %
  table [col sep=comma]{twovsnebx.csv};
\addplot [color=black,solid]
  table[row sep=crcr]{%
0.1	0.1\\
0.15	0.15\\
0.2	0.2\\
0.25	0.25\\
0.3	0.3\\
0.35	0.35\\
0.4	0.4\\
0.45	0.45\\
0.5	0.5\\
0.55	0.55\\
0.6	0.6\\
0.65	0.65\\
0.7	0.7\\
0.75	0.75\\
0.8	0.8\\
0.85	0.85\\
0.9	0.9\\
0.95	0.95\\
1	1\\
};
\legend{NEB,NEBX}
\end{axis}
\end{tikzpicture}%
  \caption{Fit of impulse response coefficients of $G_{31}$ for SMPE compared with NEB and NEBX
respectively for 100 Monte Carlo simulations. The black line represents $y = x$,
\ie when SMPE performs equally good as NEB and NEBX.}
   \label{fig:fit-smpe-vs-neb1}
\end{figure}

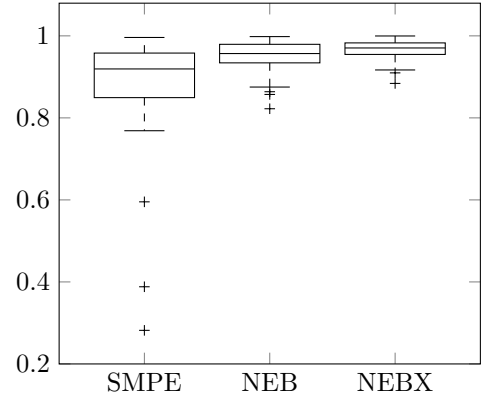
\begin{figure}[!ht]
    \centering
  \begin{tikzpicture}
	\begin{axis}[
			width=0.8\figurewidth,
			height=0.8\figureheight,
			at={(0\figurewidth,0\figureheight)},
			scale only axis,
			boxplot/draw direction=y,
			xtick={1,2,3},
			ymin = 0.2,
			xticklabels={SMPE, NEB, NEBX},]
		\addplot[boxplot,mark=+] table[y index=0]{smpenl2.csv};
		\addplot[boxplot,mark=+] table[y index=0]{nebnl2.csv};
		\addplot[boxplot,mark=+] table[y index=0]{nebxnl2.csv};
	\end{axis}
\end{tikzpicture}
  \caption{Box plot of the fit of the impulse response of $G_{32}$
          obtained by the methods SMPE, NEB and NEBX respectively.}
   \label{fig:boxplot-G32}
\end{figure}
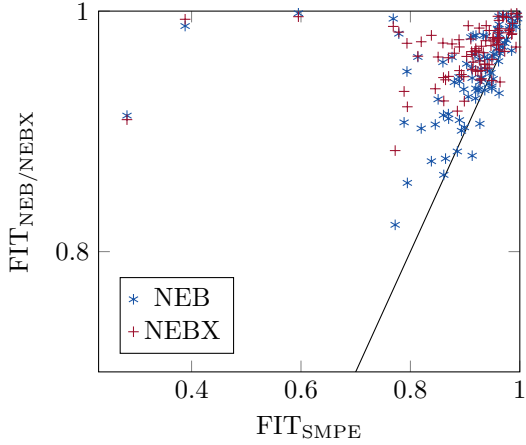
\begin{figure}[!ht]
    \centering
%

\begin{tikzpicture}

\begin{axis}[%
width=0.8\figurewidth,
height=0.8\figureheight,
at={(0\figurewidth,0\figureheight)},
scale only axis,
xmin=0.23,
xmax=1,
xtick={0.4, 0.6, 0.8, 1},
xlabel={$\text{FIT}_{\text{SMPE}}$},
ymin=0.7,
ymax=1,
ytick={0.4, 0.6, 0.8, 1},
ylabel={$\text{FIT}_{\text{NEB/NEBX}}$},
axis background/.style={fill=white},
legend style={at={(0.05,0.05)},
		anchor=south west}
]
\addplot[only marks,color=blue,mark=asterisk,mark size=2] %
  table [col sep=comma]{twovsneb2.csv};
\addplot[only marks,color=red,mark=+,mark size=2] %
  table [col sep=comma]{twovsnebx2.csv};
\addplot [color=black,solid]
  table[row sep=crcr]{%
0.1	0.1\\
0.15	0.15\\
0.2	0.2\\
0.25	0.25\\
0.3	0.3\\
0.35	0.35\\
0.4	0.4\\
0.45	0.45\\
0.5	0.5\\
0.55	0.55\\
0.6	0.6\\
0.65	0.65\\
0.7	0.7\\
0.75	0.75\\
0.8	0.8\\
0.85	0.85\\
0.9	0.9\\
0.95	0.95\\
1	1\\
};
\legend{NEB,NEBX}
\end{axis}
\end{tikzpicture}%
  \caption{Fit of impulse response coefficients of $G_{32}$ for SMPE compared
  with NEB and NEBX respectively for 100 Monte Carlo simulations. The black line
  represents $y = x$, \ie when SMPE performs equally good as NEB and NEBX.
  Note the scaling of the x-axis of this figure.}
   \label{fig:fit-smpe-vs-neb2}
\end{figure}

\section{Conclusion}
In this paper, we have addressed the identification of a module in dynamic networks with known topology.
The problem is cast as the identification of a set of systems in series
connection.
The second system corresponds to the target module, while the first
represents the dynamic relation between exogenous signals and the input and the
target module. This system is modeled following a Bayesian kernel-based approach,
which enables the identification of the target module using empirical Bayes arguments.
In particular, the target module is estimated using a marginal likelihood
criterion, whose solution is obtained by a novel iterative scheme designed through the ECM algorithm.
The method is extended to incorporate measurements downstream of the target
module, which numerical experiments suggest increases performance.

\appendix
\section{Appendix}
\subsubsection*{Proof of Lemma \ref{th:Q_function}} \label{sec:proof_lemma_Q}
From Bayes' rule it follows that
$$
\log p(z,s_{11};\,\hat \eta^{(k)}) = \log p(z|s_{11},\;\,\hat \eta^{(k)}) + \log p(s_{11};\,\hat \eta^{(k)})\,,
\IEEEeqnarraynumspace
$$
with (neglecting constant terms)
\begin{align*}
\log p(z|s_{11},\;\,\eta) & \propto  -\frac{1}{2} \log\det  \Sigma_e -\frac{1}{2} \|z - W_\theta s_{11} \|^2_{ \Sigma_e^{-1}} \nonumber \\
\log p(s_{11};\,\eta) & \propto -\frac{1}{2} \log\det \lambda K_{\beta}  -\frac{1}{2} s_{11}^T (\lambda K_{\beta})^{-1} s_{11}\,.
\end{align*}
Now we have to take the expectation w.r.t. the posterior $p(s_{11}|\tilde w_2;\,\hat \eta^{(k)})$. Developing the second term in the first equation above and recalling that
$$
\mathrm{E}_{p(s_{11}|\tilde w_2;\,\hat \eta^{(k)})}[s_{11}^T A s_{11}]  =  \tr{A \hat{S}_{11}^{(k)}}\,,
$$
the statement of the lemma readily follows.

\subsubsection*{Proof of Proposition \ref{th:update_sigma_theta}} \label{sec:proof_prop_sigma_theta}
In \eqref{eq:Q_0}, fix $\Sigma_e$ to the value $\hat \Sigma_e^{(k)}$
(computed inserting $\sigma^{2(k)}_1$ and $\sigma^{2(k)}_2$). We obtain the
$\theta$-dependent terms \eqref{eq:theta-dep1} and \eqref{eq:theta-dep2} (after multiplying by a factor $-2$),
\begin{figure*}
\begin{IEEEeqnarray}{rClCl}
-2 z^T \left(\hat \Sigma_e^{(k)}\right)^{(-1)} W_\theta \hat{s}_{11}^{(k)}
  & = & -\frac{2}{\sigma^{2(k)}_2} y^T G_\theta R_1 \hat{s}_{11}^{(k)} + k_1
    & = & -\frac{2}{\sigma^{2(k)}_2} y^T \T_N(R_1 \hat{s}_{11}^{(k)})g_\theta + k_1 \label{eq:theta-dep1} \\
\tr{ W_{\theta}^T \left( \Sigma_e^{(k)} \right)^{-1} W_{\theta}\hat S^{(k)}_{11} }
  & = & \frac{1}{\sigma^{2(k)}_2} \tr{ G_{\theta} R_1 \hat S_{11}^{(k)}  R_1^T G_{\theta}^T }  + k_2
    & = & \frac{1}{\sigma^{2(k)}_2} \vect(G_{\theta})^T (R_1 \hat S_{11}^{(k)} R_1^T \otimes I_N)\vect(G_{\theta}) + k_2 \nonumber \\
  & = &\frac{1}{\sigma^{2(k)}_2} g_{\theta}^T D^T (R_1 \hat S_{11}^{(k)}  R_1^T \otimes I_N) D g_{\theta}
    & + & k_2  \,, \label{eq:theta-dep2}
\end{IEEEeqnarray}
\end{figure*}
where $k_1$ and $k_2$ contain terms independent of $\theta$.
Recalling the definitions of $\hat A^{(k)}$ and $\hat b^{(k)}$, \eqref{eq:sol_theta} readily follows.

Now, let $\theta$ be fixed at the value $\hat \theta^{(k+1)}$.
The function \eqref{eq:Q_0} can be rewritten as \eqref{eq:Q_0_rewritten} (after multiplying by a factor $-2$).
\begin{figure*}
\begin{IEEEeqnarray}{rCl} \label{eq:Q_0_rewritten}
{Q}_0^{(k)}(\sigma^2_1,\,\sigma^2_2,\,\hat \theta^{(k+1)})
  & = & N(\log \sigma^2_1 + \log \sigma^2_2)
    + \frac{\norm{\tilde{w}_1}_2^2}{\sigma^2_1} 
   + \frac{\norm{\tilde{w}_2}_2^2}{\sigma^2_2}
   - \frac{2\tilde{w}_1^T}{\sigma^2_1} R_1 \hat s_{11}^{(k)}
   - \frac{2\tilde{w}_2^T}{\sigma^2_2}G_{\hat \theta^{(k+1)}} R_1 \hat s_{11}^{(k)} \nonumber \\
  &&  + \frac{1}{\sigma^2_1} \tr{ R_1^TR_1 \hat S^{(k)}_{11}}
   + \frac{1}{\sigma^2_2} \tr{ R_1^T G_{\hat{\theta}^{(k+1)}}^T G_{\hat{\theta}^{(k+1)}} R_1 \hat S_{11}^{(k)}}
\end{IEEEeqnarray}
\end{figure*}
The results \eqref{eq:sol_variances} follow by minimizing \eqref{eq:Q_0_rewritten} with respect to $\sigma_1^2$ and $\sigma_2^2$.
Differentiating w.r.t. $\sigma^2_1$ and $\sigma^2_2$ and calculating the zeros.

\subsection*{Proof of Proposition \ref{thm:E-step-gibbs}}
Using Bayes' rule we can decompose the complete likelihood as
\begin{IEEEeqnarray*}{rCl}
\log p(z, s_{11}, f; \eta)
	&=& \log p(z|s_{11},f; \eta) \\
	& &+ \log p(s_{11};\eta) + \log p(f;\eta) \, ,
\end{IEEEeqnarray*}
and we will analyze each term in turn.
First, note that
\begin{IEEEeqnarray*}{rCl}
-2\log p(s_{11}|\eta)
	&=&  \log \det \lambda_s K_{\beta_s} + s_{11}^T (\lambda_s
    	K_{\beta_s})^{-1}s_{11} \\
	&=& \log \det \lambda_s K_{\beta_s} + \tr{ (\lambda_s
    	K_{\beta_s})^{-1}s_{11}s_{11}^T } \label{eq:s|eta}
\end{IEEEeqnarray*}
Replacing $s_{11}s_{11}^T$ with its sample estimate yields the first term in \eqref{eq:Q-gibbs}.
Similarly,
\begin{IEEEeqnarray*}{rCl}
-2\log p(f|\eta) &=& \log \det \lambda_f K_{\beta_f}  + \tr{ (\lambda_f K_{\beta_f})^{-1}ff^T}.
\label{eq:f|eta}
\end{IEEEeqnarray*}
Replacing $ff^T$ with its sample estimate yields the second term in \eqref{eq:Q-gibbs}.
Finally,
\begin{IEEEeqnarray}{rCl}
-2 \log p(z|t,s_{11}; \eta) &=&  \log \det  \Sigma
 + (z-\hat z)^T \Sigma^{-1} (z-\hat z) \, , \nonumber
\\
\label{eq:z|eta}
\end{IEEEeqnarray}
with
\begin{IEEEeqnarray*}{rCl}
\hat z & \defeq & \begin{bmatrix}
Rs \\ G_\theta Rs \\ G_\theta Rv
\end{bmatrix} \, .
\end{IEEEeqnarray*}
The first term of \eqref{eq:z|eta} is $N$ times the sum of the logarithms of the noise variances squared.
The second term of \eqref{eq:z|eta} decomposes into a sum of the (weighted) error of each signal.
Then, the first weighted error is given by
\begin{IEEEeqnarray*}{rCl}
\sigma_1^2 \norm{\tilde w_1 - Rs}^2_2 &=&
  \norm{\tilde w_1}^2_2 - 2 \tilde w_1^T R s  +  \tr{R  s s^TR^T} \, .
\end{IEEEeqnarray*}
Replacing $s$ and $ss^T$ with their respective estimates gives the third term in \eqref{eq:Q-gibbs}, with the corresponding noise variance term of \eqref{eq:z|eta} added. Similar calculations on the remaining two weighted errors in \eqref{eq:z|eta} gives the last two terms in \eqref{eq:Q-gibbs}. This concludes the proof.

\bibliographystyle{dcu}
\bibliography{bayesian}

\end{document}